%% file: coldstart.tex
\documentclass[mnsc,nonblindrev]{informs3}

\OneAndAHalfSpacedXI % current default line spacing
\usepackage{natbib}
 \bibpunct[, ]{(}{)}{,}{a}{}{,}%
 \def\bibfont{\small}%
\TheoremsNumberedThrough     % Preferred (Theorem 1, Lemma 1, Theorem 2)
\ECRepeatTheorems

\EquationsNumberedThrough    % Default: (1), (2), ...
\MANUSCRIPTNO{}

\begin{document}
%%%%%%%%%%%%%%%%

%\setcounter{page}{1}
%\DOI{10.1287/xxxx.0000.0000}%

\RUNAUTHOR{Holtz et al.}

\RUNTITLE{The Engagement-Diversity Connection}

% \TITLE{Bundling Information Goods of Decreasing Value}
\TITLE{The Engagement-Diversity Connection: Evidence from a Field Experiment on Spotify}
% Optimizing solely for engagement can have negative repercussions for diversity
% Impact of optimizing for engagement on diversity: evidence from a field experiment on Spotify
% 

\ARTICLEAUTHORS{%
\AUTHOR{David Holtz}
\AFF{MIT Sloan School of Management, Cambridge, MA 02139, \EMAIL{dholtz@mit.edu}} %, \URL{}}
\AUTHOR{Benjamin Carterette, Praveen Chandar, Zahra Nazari, Henriette Cramer}
\AFF{Spotify, Inc., New York, NY 10007, \EMAIL{benjaminc@spotify.com}, \EMAIL{praveenrchandar@spotify.com}, \EMAIL{zahran@spotify.com}, \EMAIL{henriette@spotify.com}}
\AUTHOR{Sinan Aral}
\AFF{MIT Sloan School of Management, Cambridge, MA 02139, \EMAIL{sinan@mit.edu}}
} % end of the block

\ABSTRACT{%
It remains unknown whether personalized recommendations increase or decrease the diversity of content people consume. We present results from a randomized field experiment on Spotify testing the effect of personalized recommendations on consumption diversity. In the experiment, both control and treatment users were given podcast recommendations, with the sole aim of increasing podcast consumption. Treatment users' recommendations were personalized based on their music listening history, whereas control users were recommended popular podcasts among users in their demographic group. We find that, on average, the treatment increased podcast streams by 28.90\%. However, the treatment also decreased the average individual-level diversity of podcast streams by 11.51\%, and increased the aggregate diversity of podcast streams by 5.96\%, indicating that personalized recommendations have the potential to create patterns of consumption that are homogenous within and diverse across users, a pattern reflecting Balkanization. Our results provide evidence of an ``engagement-diversity trade-off" when recommendations are optimized solely to drive consumption: while personalized recommendations increase user engagement, they also affect the diversity of consumed content. This shift in consumption diversity can affect user retention and lifetime value, and impact the optimal strategy for content producers. We also observe evidence that our treatment affected streams from sections of Spotify's app not directly affected by the experiment, suggesting that exposure to personalized recommendations can affect the content that users consume organically. We believe these findings highlight the need for academics and practitioners to continue investing in personalization methods that explicitly take into account the diversity of content recommended.
}%
%\KEYWORDS{deterministic inventory theory; infinite linear programming duality;
%  existence of optimal policies; semi-Markov decision process; cyclic schedule}

%\KEYWORDS{%butter, margarine, silliness
%} 
%\HISTORY{%This paper was first submitted on April 12, 1922 and has been with the authors for 83 years for 65 revisions.
%}

\maketitle

\section{Introduction}

Recommender systems and algorithmic content curation play an increasingly large role in people's lives. For instance, algorithmic recommendations influence the news and entertainment that we consume, the products that we purchase, and the people with whom we develop romantic relationships. Collaborative filtering recommendation systems drive 35\% of product choices on Amazon \citep{lamere2008project} and 60\% of consumption choices on Netflix \citep{thompson2008if}. However, despite recommender systems' increasing ubiquity, the ways in which they impact the \textit{types} of choices we make are still not well understood. While some scholars have speculated that recommender systems will lead to ``filter bubbles" \citep{sunstein2001republic, pariser2011filter}, others hypothesize that recommender systems will homogenize user consumption, leading to the ``rich getting richer" \citep{negroponte1996being, van2005global, salganik2006experimental}. In this paper, we analyze a large scale field experiment conducted on Spotify, one of the world's leading streaming platforms. During the experiment, both treatment and control users were recommended podcasts with the sole aim of increasing podcast consumption; while control users were recommended podcasts popular amongst those in their demographic group, treatment users were provided fully personalized recommendations based on their existing music listening history. We measure the impact of more personalized content recommendations on user engagement, as well as individual-level and aggregate podcast category diversity.

We find that the recommender system in the experiment increased the average number of podcasts streamed per user by 28.90\% relative to the less-personalized, popularity-based recommendation strategy. We also test for the impact of the algorithm on user-level podcast category diversity, as measured through the Shannon entropy \citep{shannon1948mathematical, teachman1980analysis}, and aggregate podcast category diversity, as measured through a quantity that we call ``intragroup diversity" \citep{aral2016unpacking}.\footnote{While we quantify diversity using these particular measures, there is a large academic literature discussing different approaches to measuring diversity. See, for instance, \citet{mitchell2020diversity}.} We find that the more personalized algorithm \textit{decreased} individual-level diversity,  but \textit{increased} intragroup diversity. These results indicate that recommender systems and personalization algorithms have the capacity to push individual users into homogeneous consumption patterns that are increasingly dissimilar from those of their peers. While the effects of the treatment are largest for streams originating from the section of Spotify's app where personalized recommendations are delivered, we observe evidence that the treatment also affected streams originating from other parts of the app. This suggests that exposure to personalized recommendations can also affect the diversity of content that users engage with organically. 

In aggregate, our findings highlight the potential for recommender systems to create an ``engagement-diversity trade-off" for firms when recommendations are optimized solely to drive consumption; while algorithmic recommendations can increase user engagement, they can also homogenize individual users' consumption and Balkanize user content consumption. This shift in consumption diversity can negatively impact user churn rates and lifetime values \citep{anderson2019}, and can impact the optimal strategy for content creators, including platforms that create original content (such as Spotify or Netflix). It is possible that our findings also extend to cases where diversity is measured with respect to the ideological slant or extremity of content. If so, the ``engagement-diversity tradeoff" suggests that recommender systems that increase engagement/consumption can also create costs for firms, due to the high level of public scrutiny given to personalized recommendations, and impact public discourse on platforms through the creation of ``filter bubbles." In light of our results, we believe it is worthwhile for academics and practitioners to continue developing personalization techniques that explicitly take into account the diversity of content recommended to users \citep{marler2004survey, castells2015novelty, lacerda2017multi}.

%The fact that our treatment has much smaller effects on consumption coming from other parts of Spotify's app, and does not lead to longer-term effects also suggests that sustained exposure to algorithmic recommendations may play a role in the creation of online echo chambers and radicalization pathways.

Our work contributes to an emerging research literature that uses randomized field experiments to measure the impact of recommender systems on the \textit{diversity} of content that users consume \citep{claussen2019editor, lee2019recommender}. Notably, our results indicate that recommender systems can decrease individual-level diversity while increasing aggregate diversity, while \citet{lee2019recommender} find that the introduction of algorithmic recommendations had a neutral-to-positive effect on individual-level diversity while decreasing aggregate diversity. We believe this contrast is due to differences in how the two studies quantify diversity, as well as differences in the recommendation algorithms being used and the research settings in which the two experiments are conducted. These conflicting results highlight the importance of measuring the impact of many different recommendation algorithms in a variety of settings, and the need to develop a number of different methodological approaches for measuring the impact of recommendation systems on consumption diversity.

%The rest of this paper proceeds as follows. Section \ref{sec:literature} reviews the related literature. Section \ref{sec:setting} describes our research setting, podcasts on Spotify, and Section \ref{sec:experiment} describes the design of our experiment. We present our results in Section \ref{sec:results}, and discuss our findings in Section \ref{sec:discussion}. In Section \ref{sec:conclusion}, we conclude.

\section{Related Literature} \label{sec:literature}

This paper contributes to a growing body of literature that focuses on the economic and societal impacts of recommender systems, which use product metadata as well as implicit and explicit user feedback to generate personalized product recommendations to users \citep{resnick1997recommender, adomavicius2005toward}. Early research established that online recommendations impact consumer product choices \citep{senecal2004influence}, and that recommender systems in particular often lead to increased engagement and/or purchases \citep{das2007google, freyne2009increasing, de2010technology, zhou2010impact, oestreicher2012visible, sharma2018split}. However, there is no clear consensus on the impact that recommender systems have on the \textit{diversity} of items that users consume. 

Building on the work of \cite{brynjolfsson2011goodbye}, a series of papers have attempted to quantify, through models, simulations, observational analysis, and natural experiments, the effect of recommender systems on sales diversity \citep{fleder2009blockbuster, wu2011recommendation, oestreicher2012recommendation, jannach2013recommenders, hosanagar2013will, nguyen2014exploring, hervas2015recommended}. Most, but not all, of these papers measure changes in sales diversity by looking at differences in the Lorenz curve corresponding to product consumption or sales. 
%Although the findings of any particular study are dependent on the algorithm being studied and the setting in which it is deployed, 
Many of these studies argue that recommender systems make individual consumption more diverse, while \textit{decreasing} aggregate consumption diversity. To provide some intuition for how this might occur, imagine a platform with four users and four pieces of content: $A$, $B$, $C$, and $D$. A recommender system could shift users' consumption vectors from ($A$), ($B$), ($C$), ($D$) to ($AB$), ($AB$), ($AB$), ($AB$). While each individual users' consumption is less concentrated, aggregate consumption is more concentrated.

A separate stream of research has focused on the impact that recommender systems have on the \textit{types} of content that people consume, and the resultant societal impacts. While some papers in this research stream argue that algorithms lead to increased ideological segregation \citep{flaxman2016filter, tufekci2018youtube, ribeiro2019auditing}, others find that users' tendency to engage with content that agrees with their ideological preferences is driven by user choice, as opposed to algorithms \citep{gentzkow2006media, bakshy2015exposure}. In this paper, we focus on diversity with respect to podcast categories, rather than ideological affiliation. However, both types of diversity characterize the \textit{type} of content that users consume, and it is possible that the application of our analytical framework to data with ideological labels would produce similar results.

Three recent papers closely related to our research are \citet{claussen2019editor, lee2019recommender} and \citet{anderson2019}. Both \citet{claussen2019editor} and \citet{lee2019recommender} conduct online field experiments that measure the effect of personalized recommendations on consumption diversity. \citet{claussen2019editor} find that personalization \textit{decreases} individual-level diversity, and that this decrease in consumption diversity spills over to non-personalized sections of the website they study. On the other hand, \citet{lee2019recommender} find that the introduction of a recommender system had a neutral-to-positive effect on individual-level diversity, but decreased aggregate diversity. \citet{anderson2019} use observational data from Spotify to study the relationship between personalization and listening diversity. They find that user-driven listening is more diverse than algorithmic listening, and that users who become more diverse over time do so by shifting away from algorithmic listening. Importantly, they also find that users with more diverse listening habits are less likely to leave the platform and are more likely to eventually become paid subscribers. 

Our research builds on the existing literature in multiple ways. First, in contrast to many observational studies of recommender systems, we analyze data from a randomized field experiment, which enables us to credibly estimate the causal effect of personalized recommendations on content consumption. Second, whereas most recommender systems research in economics and management has focused on measuring changes in sales diversity, we focus on measures of diversity that take into account the \textit{types} of content that users consume, as measured through podcast category tags on Spotify. Finally, we study the impact of a novel algorithm (which predicts podcast affinity based on a user's music listening history) in a novel setting (podcast recommendations on Spotify). Given that the impact of recommender systems can depend on a wide range of factors, including but not limited to the type of data used for training \citep{lin2015demand}, the algorithm used to generate recommendations \citep{wu2011recommendation, jannach2013recommenders}, and the setting in which the recommender is deployed, it is important for researchers to continue studying the impact of many different recommendation algorithms in a number of different settings.

\section{Research Setting} \label{sec:setting}

\subsection{Spotify}

The setting for our study is Spotify, one of the world's leading streaming platforms. Spotify was founded in 2006, and as of December 2019, has 271 million monthly active users and 124 million paying subscribers.\footnote{\url{https://s22.q4cdn.com/540910603/files/doc_financials/2019/q4/Shareholder-Letter-Q4-2019.pdf}} Although Spotify launched as a music streaming platform, in 2015 the company expanded its offerings to include videos and, more importantly for this study, podcasts.\footnote{\url{https://www.fastcompany.com/3046504/spotify-launches-podcasts-video-and-context-based-listening}} Podcasts are an increasingly popular type of content to stream online, and represent an important new vertical for Spotify. According to Edison Research, 51\% of the U.S. population has listened to at least one podcast, and 32\% listens to podcasts on a monthly basis. Among monthly podcast listeners, 43\% have listened to a podcast on Spotify \citep{edison2019}.

Spotify users on mobile are able to access three different sections of the Spotify app via a navigation bar that runs along the bottom of the phone screen: ``Your Library," ``Search," and ``Home." The ``Home" section of the app is most relevant to our research. It presents the user with a ranked set of ``shelves," each of which contains a ranked set of ``cards." Shelves correspond to different types of content, such as ``content a user was recently listening to," or "music from a particular genre." Each card is essentially a link to a piece of content (e.g., a playlist or Spotify artist page). Shelves on home, and the cards within each shelf, are ordered by a combination of machine learning algorithms and human editors.\footnote{Details of Spotify's approach to ranking home content can be found in \citet{mcinerney2018explore}.} A screenshot of the ``Home" section of the Spotify app on iOS can be seen in Figure \ref{fig:home_layout}. 

%The ``Your Library" section of the app allows a user to access albums, playlists and albums that they have previously saved, as well as podcasts they have previously followed and podcast episodes they have previously downloaded. The ``Search" section of the app allows users to search Spotify's content library for specific pieces of content. 

In this paper, we will analyze changes to the number of podcasts users stream, as well as the \textit{types} of podcasts users stream. Each podcast stream has a ``referrer" field associated with it, which indicates which part of the Spotify app the stream originated from. This field allows us to differentiate between streams that originated on the ``Home" section of the app (where the experiment introduced variation in recommended podcasts) and streams that originated from other sections of the app (where the experiment did not introduce any variation).

\subsection{Podcast categorization}

At the time of the experiment, there were thirteen podcast category tags that could be associated with a podcast on Spotify: ``Arts \& Entertainment," ``Business \& Technology," ``Comedy," ``Educational," ``Games," ``Kids \& Family," ``Lifestyle \& Health," ``Music," ``News \& Politics," ``Society \& Culture," ``Sports \& Recreation," ``Stories," and ``True Crime." In our dataset, there are as many as ten podcast category tags associated with any particular podcast. However, 68.23\% of podcasts have only one category associated with them, and 97.50\% of podcasts have three or fewer podcasts associated with them. We use the category tags associated with each user's podcast streams to quantify changes in the diversity of their podcast consumption. For streams of podcasts that have multiple category tags, we divide the stream evenly across each of the podcast's associated categories. A more detailed description of podcast categorization on Spotify can be found in Appendix \ref{sec:podcast_categories}.

\section{Experiment Design} \label{sec:experiment}

We analyze data from an experiment conducted on a sample of 852,937 premium Spotify users across seventeen countries\footnote{The experiment was conducted on users located in AR, AU, BR, CA, CL, CO, DE, DK, ES, FR, GB, IT, MX, NL, NO, RS, and US.} between April 18, 2019 and May 2, 2019 as part of a product rollout. In order to be eligible for the experiment, a user needed to have never streamed or followed a podcast on Spotify, and to have visited the ``Home" section of the Spotify app during the experiment. Users were assigned to treatment arms using a ``bucket randomization" procedure. That randomization procedure, along with the balance of observable characteristics between the treatment and control groups, is described in Appendix \ref{sec:randomization}.

\begin{figure}[!htbp]
\begin{center}
\includegraphics[height=4in]{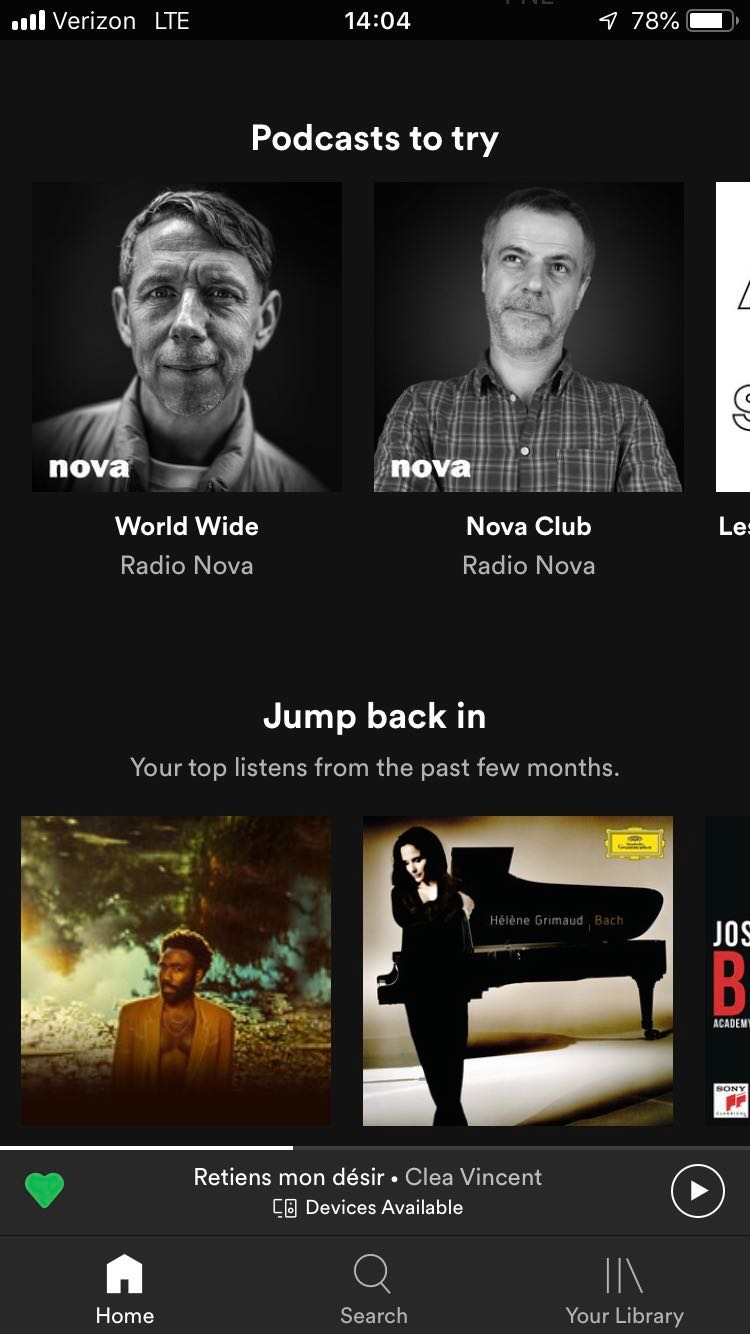}
\caption{A screenshot of the ``Podcasts to try" shelf on the Spotify iOS app. During the experiment, this shelf was fixed in the second slot on the ``Home" section of the Spotify app.}
\label{fig:home_layout}
\end{center}
\end{figure}

Users in both the treatment and control were exposed to a shelf in the ``Home" section of the Spotify mobile app labeled ``Podcasts to Try," which was anchored in the second highest slot in the ``Home" section. For users in the treatment, the ``Podcasts to Try" shelf was populated with 10 recommendations generated by a neural network model that predicted the podcasts a user would follow based on their music listening history and demographic information.\footnote{For a more detailed description of the neural network model, we refer the reader to \cite{nazari2019recommending}.} For users in the control, the ``Podcasts to Try" shelf was populated with the 10 most popular podcasts among users who shared the focal user's self-reported gender, age bucket, and country.\footnote{Age buckets are defined as follows: 18-24, 25-29, 30-34, 35-44, 45-54, 55+.} Both the machine learned recommendations and the demographic-based recommendations were determined using pre-treatment data, and were not updated over the course of the experiment. For users in both treatment arms, the ``Podcasts to Try" shelf was hidden once the user had streamed or followed any podcast on Spotify. Figure \ref{fig:home_layout} shows a screenshot of the ``Podcasts to try" shelf on iOS. The shelf's UI was consistent across the control and treatment groups; the only thing exogenously varied was the set of podcasts populating the shelf.

\section{Results} \label{sec:results}

In this section, we present the experiment results. 
% We first measure the impact of the treatment on the volume of content consumed. We proceed to measure the impact of the treatment on the \textit{diversity} of content consumed. Finally, we then test for differences in the effect of the treatment across different streaming source within the app. 
We report the effects of the treatment on podcast streams, however, the effects of the treatment on podcast follows are extremely similar, and can be found in Appendix \ref{sec:app_follows}.

\subsection{Effect on podcast consumption}

We first study the impact of algorithmic podcast recommendations on the average number of podcast streams per user during the experiment.\footnote{ The long-term effects of the experiment are reported in Appendix \ref{sec:long_term_effects}.}\footnote{The effect of the treatment on the number of users streaming at least one podcast is reported in Appendix \ref{sec:podcast_streamers}.} We estimate the effect of the treatment by estimating the following model:

\begin{equation} \label{eq:baseline_model}
y_i = \alpha + \beta T_i + \delta X_i + \epsilon_i,
\end{equation}

\noindent where $y_i$ is the outcome of interest for user $i$ (in this case, podcast streams), $\alpha$ is a constant, $X_i$ is a vector of user-level covariates (age bucket, self-reported gender, and account age in days), and $T_i$ is user $i$'s treatment assignment. Standard errors are clustered at the user treatment assignment bucket-level.

\begin{figure}[!htbp]
\begin{center}
\includegraphics{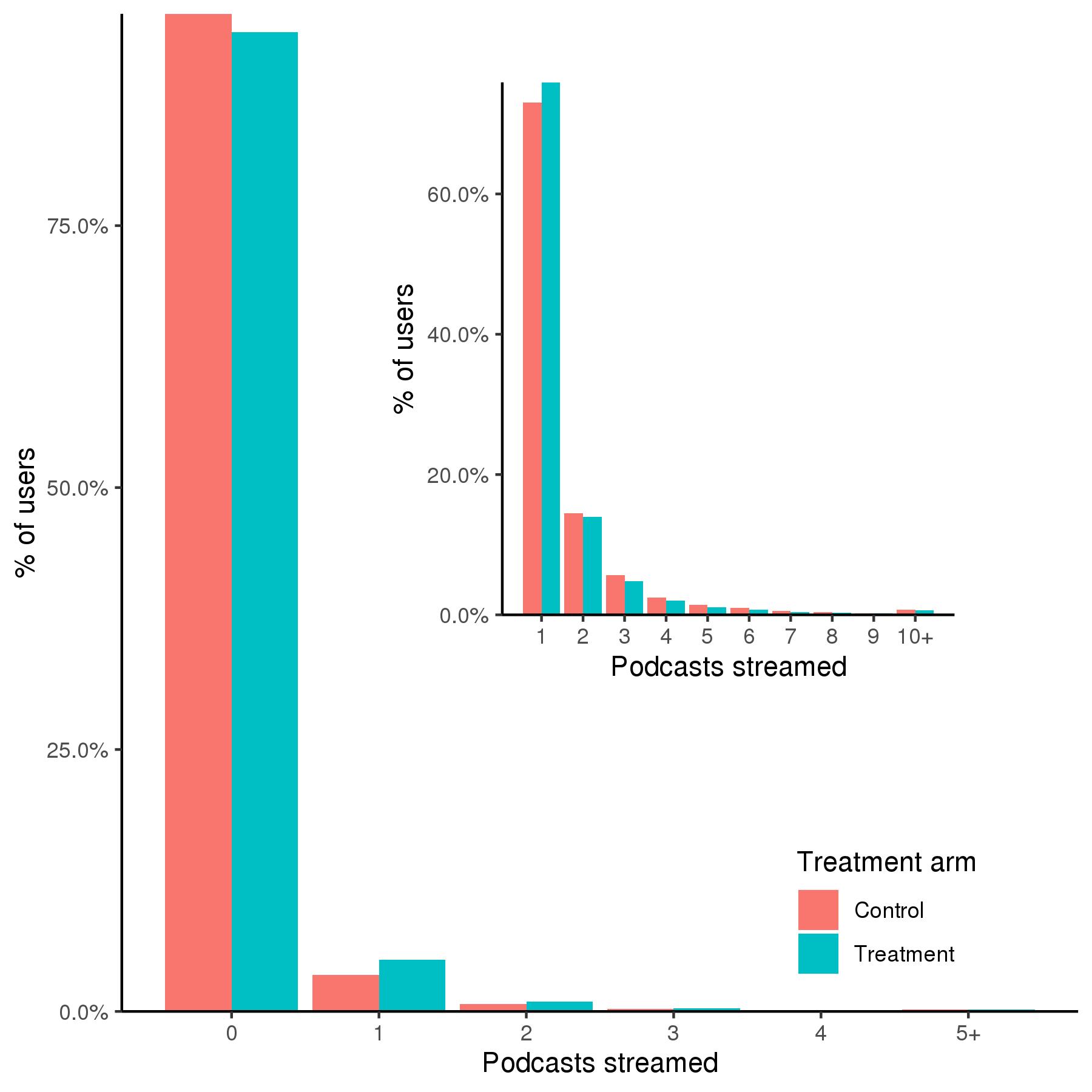}
\caption{The distribution of of podcasts streamed in both treatment arms. Inset plot shows the distribution of podcasts streamed in both treatment arms conditional on streaming at least one podcast.}
\label{fig:inset_streams_per_user}
\end{center}
\end{figure}

Figure \ref{fig:inset_streams_per_user} shows the distribution of podcast streams per user during the experiment in both treatment arms, both overall and conditional on the user streaming at least one podcast during the experiment. Table \ref{tab:shows_played_model} reports the estimated effect of the treatment on podcast streams per user during the experiment, both with and without controlling for user-level covariates. We find that the treatment increased the number of podcast streams per user by 28.90\% ($\pm$ 3.81\%). This large treatment effect indicates that personalized podcast recommendations were extremely effective at increasing podcast consumption during the experiment.

Using the principal stratification approach detailed by \citet{frangakisPrincipalStratificationCausal2002} and \citet{ding2017principal}, we are also able to measure the extent to which this treatment effect is driven by compositional shifts, as opposed to intensity shifts.\footnote{The principal stratification methodology is detailed in Appendix \ref{sec:principal_strat}.} We estimate that on average, ``compliers" (i.e., those who would only stream at least one podcast if exposed to the treatment) streamed 1.505 (95\% CI: (1.488, 1.522)) more podcasts in the treatment, whereas ``always takers" (i.e., those who would stream at least one podcast whether in the control or treatment) streamed 0.082 (95\% CI: (0.055, 0.112)) \textit{fewer} podcasts in the treatment. In other words, the increase in podcast streaming is driven by a greater number of users streaming \textit{at least one} podcast during the experiment, as opposed to an increase in the amount of podcast streaming from those who would have streamed at least one podcast even if they had not been exposed to the treatment.

\subsection{Effect on diversity of podcast consumption}

We also measure the effect of the treatment on the diversity of content that individual users consume (henceforth referred to as ``individual-level diversity") and the diversity of content consumption across users (henceforth referred to as ``intragroup diversity").

\subsubsection{Individual-level diversity}

We quantify individual-level diversity using the Shannon entropy \citep{shannon1948mathematical}.\footnote{The Shannon entropy is also sometimes referred to as the Teachman index \citep{teachman1980analysis}.} The Shannon entropy of user $i$'s streams is defined as

\begin{equation} \label{teachman}
th_i = - \sum_{c \in C} s_{ci} \ln(s_{ci}),
\end{equation}

\noindent where $C$ is the full set of podcast categories and $s_{ci}$ is the share of user $i$'s streaming coming from category $c$. 

Note that if a user did not stream any podcasts belonging to category $c$, that podcast category's contribution to the Shannon entropy is zero. Importantly, this also means that users who did not listen to \textit{any} podcasts during the experiment have a Shannon entropy of zero. This, along with the fact that the treatment had a large, positive effect on the number of users streaming podcasts, could cause the observed effect of the treatment on Shannon entropy across all users to be positive, even if consumption conditional on streaming became less diverse.

\begin{figure}[!htbp]
\begin{center} 
\includegraphics{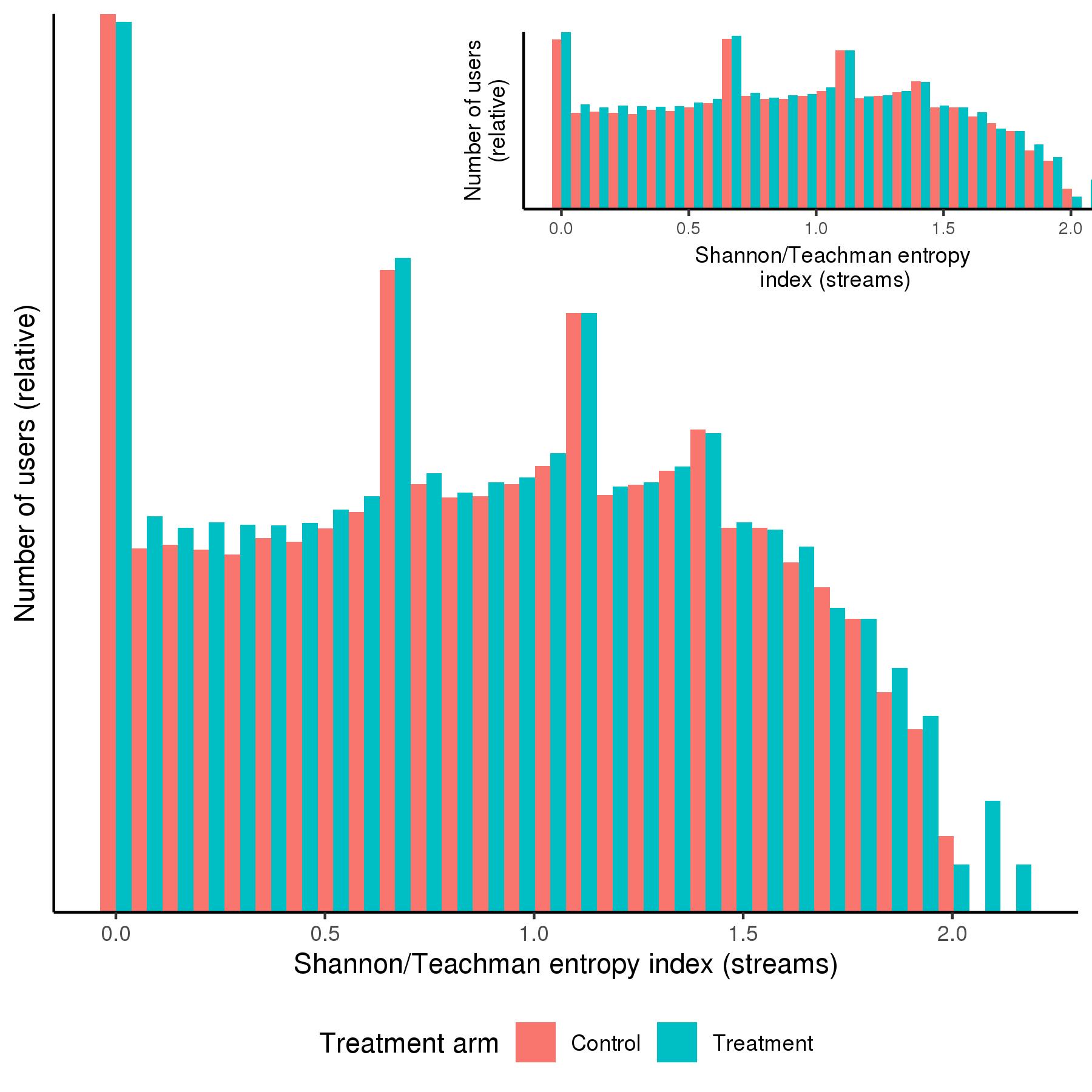}
\caption{The distribution of the user-level diversity for streams in both treatment arms. Inset plot shows the distribution of user-level diversity in both treatment arms conditional on streaming at least one podcast. y-axis values are on a log scale, and are hidden due to confidentiality concerns.}
\label{fig:inset_shows_streamed_diversity_plot}
\end{center}
\end{figure}

To account for this, we conduct our analysis on the subset of users that streamed at least one podcast during the experiment.\footnote{Analysis on the full sample can be found in Appendix \ref{sec:ind_div_all}. As expected, the measured effect of the treatment is positive.} Results of this analysis cannot be interpreted as causal user-level effects, since we are conditioning on a post-treatment variable (streaming at least one podcast). Nonetheless, these results provide some insight into the extent to which increased recommendation personalization changed individual-level diversity. Figure \ref{fig:inset_shows_streamed_diversity_plot} shows the histogram of user-level Shannon entropy for podcast streams in both treatment arms, and Table \ref{tab:individual_diversity_model_streamers_only} reports the difference in the average streaming user's Shannon entropy, both with and without controlling for user-level covariates.\footnote{Figure \ref{fig:user_level_diversity_histogram_streamers_only_by_streams} shows the distributions of user-level Shannon entropy in both treatment arms conditional on streaming a particular number of podcasts during the experiment.} We find that the average Shannon entropy of podcast streams among users who streamed at least one podcast was 11.51\% ($\pm 1.08\%$) lower in the treatment. Although this result is non-causal, we can use the principal stratification approach to estimate the causal effect of the treatment on individual-level diversity for the subset of users who are ``always takers." Consistent with the previously reported non-causal findings, we estimate that treatment decreased the average Shannon entropy for streaming ``always takers" by 0.070 (95\% CI: (0.062, 0.076)).

% CACE effect: 0.546 (95\% CI: (0.539, 0.552))

The fact that higher levels of recommendation personalization decreased the average Shannon entropy for always takers indicates that the treatment made users' podcast consumption \textit{more homogenous} with respect to podcast categories. Our analysis cannot identify to what extent this difference is driven by treatment users streaming podcasts that had fewer podcast categories associated with them. However, insofar as podcast categories accurately capture information about the topics covered in a particular show, it is reasonable to assume that a user who listened to podcasts with fewer category tags conditional on streaming a particular number of podcasts consumed less diverse content.

\subsubsection{Intragroup diversity}

We quantify intragroup diversity using a mathematical expression introduced by \citet{aral2016unpacking}:

\begin{equation} \label{intragroup_diversity}
ID = \frac{1}{n_c} \sum_{j=1}^{n_c} \left [ 1 - \cos \left ( \Gamma_{j}, \bar{\Gamma} \right ) \right ]^2,
\end{equation}

\noindent where $n_c$ is the number of users consuming at least one podcast, $\Gamma_j$ is a vector describing the fraction of user $j$'s listening belonging to each podcast category, and $\bar{\Gamma}$ is the average of $\Gamma_j$ across all users streaming at least one podcast. Intuitively, $ID$ measures the variance of all streaming users' individual-level podcast category consumption vectors. We calculate $ID$ separately for the control and treatment groups, and test for a statistically significant difference. 

We find that the treatment increased the intragroup diversity for podcast streams by 5.96\% (95\% CI: 5.45\%, 6.44\%)), from 0.710 (95\% CI: (0.708, 0.713)) in the control group to 0.753 (95\% CI: (0.751, 0.754)).\footnote{95\% confidence intervals calculated with the cluster bootstrap ($n_{boot} = 1,000$).} These results indicate that the treatment had a causal effect on the variance of podcast streamers' individual-level category consumption vectors.
In other words, not only did increased recommendation personalization push podcast streamers to consume more homogenous content, it also pushed podcast streamers to listen to content that was \textit{more} dissimilar to the content that other streamers listened to.

% The difference in intragroup diversity between the treatment and control is 0.042 (95\% CI: (0.039, 0.046)), which translates to an increase of 5.96\% (95\% CI: 5.45\%, 6.44\%)).

\subsection{Treatment effects by stream referrer}

In this subsection, we present the effects of the treatment on streams originating from different sections of the Spotify app. Because the treatment only directly affects the podcasts that are displayed on ``Home," stream referrer-level treatment effects provide insight into the extent to which exposure to personalized content recommendations impacted the types of podcasts that users sought out organically. If exposure to recommendations \textit{did} change what users sought out organically, we would expect the treatment to impact what users stream from other parts of Spotify's app, such as ``Search" and  ``Your Library." As a result, we would observe treatment effects for streams originating from both ``Home" and from non-home surfaces. On the other hand, if the treatment did not change what users sought out organically (i.e., treatment effects are entirely driven by what users consume when streaming recommended content on ``Home"), we would expect to see treatment effects for ``Home" streams, but no treatment effects for non-home streams.

%While the treatment exogenously varied the set of podcasts recommended to users in the ``Home" section of the app, it did not introduce any direct variation to the podcasts shown on any other referral surfaces, such as ``Your Library" or ``Search."

\subsubsection{Podcast consumption}

\begin{figure}[!htbp]
\begin{center} 
\includegraphics{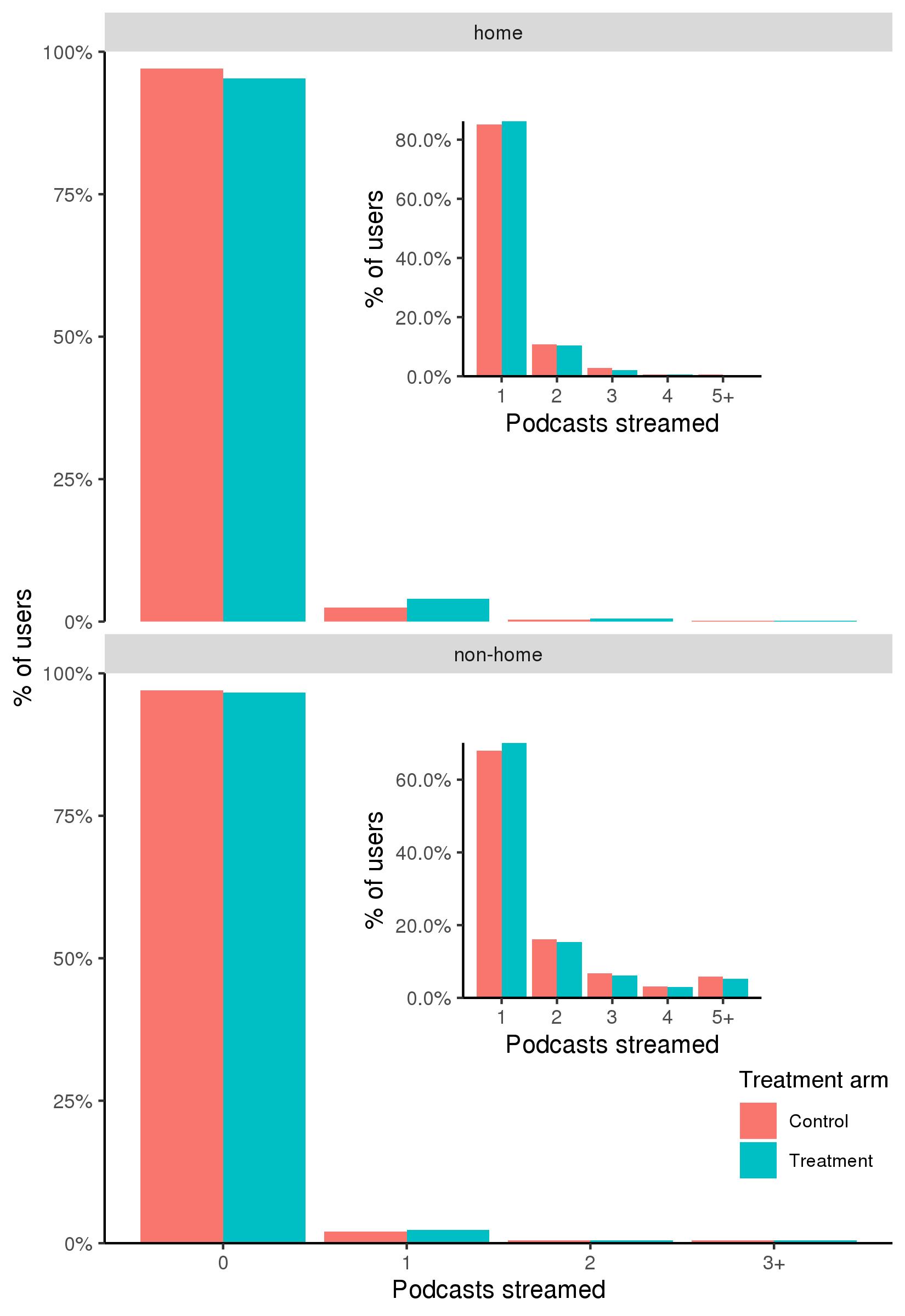}
\caption{The distribution of podcasts streamed in both treatment arms by stream referrer. Inset plots shows the distribution of podcasts streamed by referrer in both treatment arms conditional on streaming at least one podcast.}
\label{fig:streams_histogram_inset_popularity}
\end{center}
\end{figure}

%We first report the effect on the treatment on the number of podcasts that Spotify users stream from both home and non-home surfaces. 
Figure \ref{fig:streams_histogram_inset_popularity}  shows the distribution of podcast streams per user on both types of referral surfaces in both treatment arms, and Table \ref{tab:shows_played_model_referrer} reports the estimated effect of the treatment on podcast streams per user from either type of referral surface during the experiment, both with and without controlling for user-level covariates. We find that the treatment increased the average number of podcast streams per user from home by 55.75\% ($\pm 5.07\%$), and the average number of podcast streams per user from non-home surfaces by 10.47\% ($\pm 4.16\%$). 
%In other words, the treatment not only caused an increase in podcast streaming behavior from the ``Home" section of the Spotify app, but also led to a (smaller) increase in the amount of podcast streaming behavior from other sections of the app. 

\subsubsection{Individual-level diversity}

\begin{figure}[!htbp]
\begin{center} 
\includegraphics{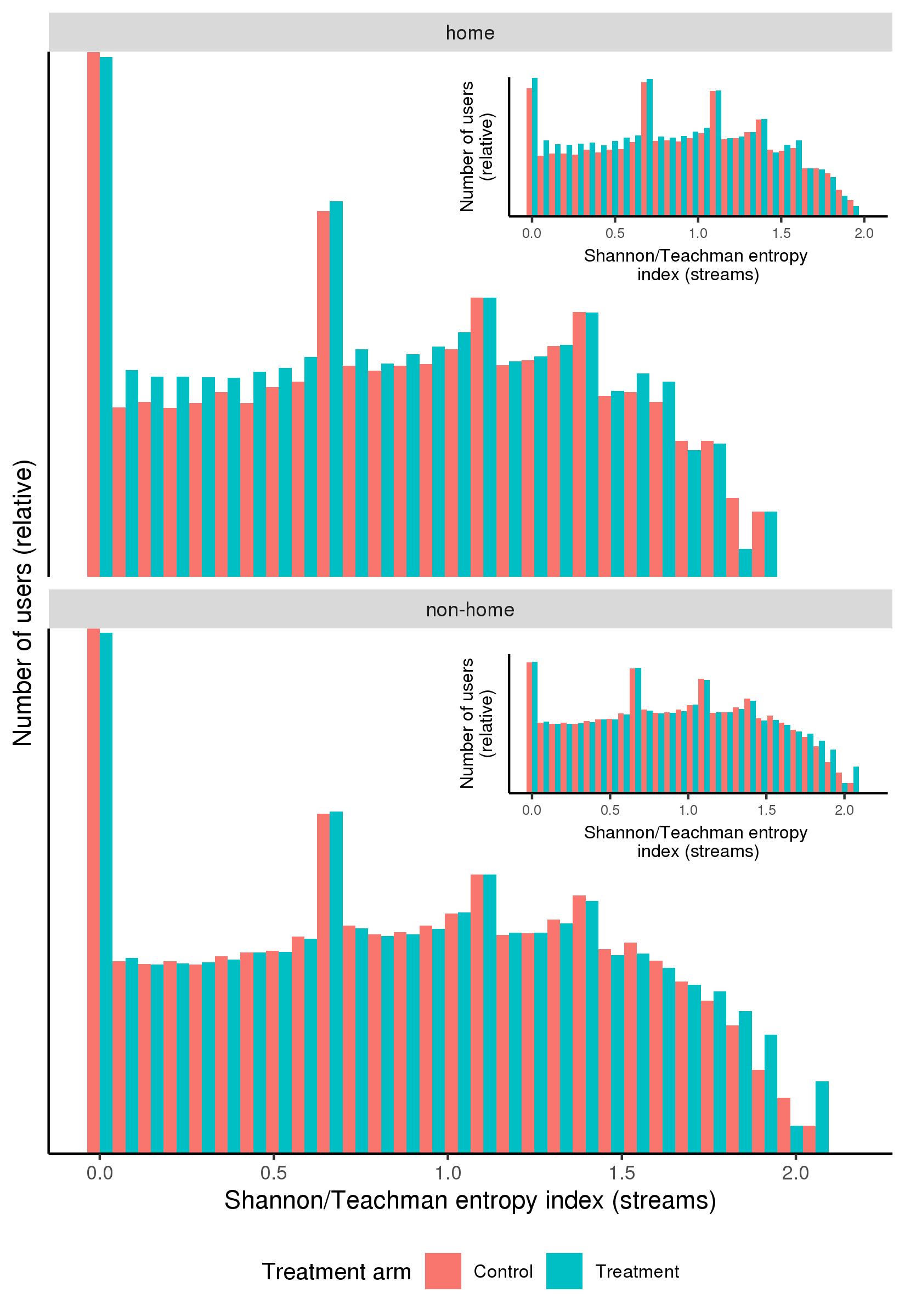}
\caption{The distribution of individual-level diversity for podcast streams in both treatment arms by stream referrer. Inset plots show the distribution of individual-level diversity by referrer in both treatment arms conditional on streaming at least one podcast. y-axis values are on a log scale, and are hidden due to confidentiality concerns.}
\label{fig:streams_diversity_histogram_inset_referrer}
\end{center}
\end{figure}

%We next report the effect of the treatment on the Shannon entropy for streams originating on home and non-home surfaces. 
Figure \ref{fig:streams_diversity_histogram_inset_referrer} shows histograms of the user-level Shannon entropy for podcast streams across both types of referral surface for users in both treatment arms, and Table \ref{tab:individual_diversity_model_streams_only_referrer} reports the differences in the average referrer-specific, user-level Shannon entropy for the subsample of users streaming at least one podcast from a given surface type during the experiment, both with and without controlling for user-level covariates. We find that for users streaming at least one podcast from ``Home," the average Shannon entropy of ``Home" streams was 17.70\% ($\pm$ 1.10\%) lower in the treatment group, and that for users streaming at least one podcast from a section other than ``Home," the average Shannon entropy of non-home streams was 3.31\% ($\pm$ 1.77\%) lower in the treatment group. 
%These results indicate that individual-level diversity in the treatment group was not only lower for streams originating from ``Home," but also for streams coming from other sections of the Spotify app.

\subsubsection{Intragroup diversity}

% Finally, we calculate the effect of the treatment on the intragroup diversity for streams originating on ``Home" and non-home referral surfaces. 
We find that on ``Home," the intragroup diversity increased by 14.04\% (95\% CI: (13.34\%, 14.66\%), from 0.654 (95\% CI: 0.650, 0.657) in the control group to 0.746 (95\% CI: (0.744, 0.748)). We find that on non-home surfaces, the intragroup diversity increased by 0.040\% (95\% CI: (-0.19\%, 0.96\%), from 0.769 (95\% CI: (0.765, 0.771)) in the control group to 0.772 (95\% CI: (0.769, 0.775)). In other words, while we do find evidence of an increase in intragroup diversity for streams originating on ``Home," we do not find statistically significant evidence of an increase in intragroup diversity for non-home streams.

%The difference in intragroup diversity between the treatment and control for home streams is 0.092 (95\% CI: (0.088, 0.095))
%The difference in intragroup diversity for non-home streams was 0.003 (95\% CI: (-0.001, 0.007))

\section{Discussion} \label{sec:discussion}

We find that personalized recommendations not only increased content consumption, but also increased the homogeneity of content consumed by individual users and increased the diversity of content consumed across users. These results suggest that an ``engagement-diversity tradeoff" can exist for firms that utilize personalization algorithms and recommendation systems to increase engagement and/or sales. This trade-off has multiple managerial implications. First, \citet{anderson2019} find that higher levels of individual-level diversity are associated with lower churn rates and higher rates of premium service subscriptions. If this relationship is causal, this would suggest that short-term increases in engagement/sales arising from the use of recommendation systems can have a neutral or even negative long-run effect on revenue. Second, the fact that recommendation systems can decrease individual-level diversity, but increase aggregate diversity may affect the optimal strategy for content creators, including platforms that produce their own original content (e.g., Spotify, Netflix). Depending on the diversity of content that users consume, content creators may find it optimal to produce large amounts of low-budget, niche content, or a small amount of high-budget content with mass appeal. Finally, in this paper, we measure the effect of increased personalization on consumption diversity measured with respect to podcast categories. However, it's possible that our analytical framework, if applied to data with ideological labels, would yield similar results. If this is the case, when the content delivered by a platform is ideological and/or extreme in nature, recommender systems that increase short term firm revenue could also create costs for firms due to the high level of public scrutiny given to personalized recommendations, and impact the nature of public discourse through the creation of ``filter bubbles."

Our results also shed light on the effect that exposure to personalized recommendations has on the types of content that users seek out organically. Although we observe stronger treatment effects on streams originating from the ``Home" section of Spotify's app, the treatment does affect the volume and individual-level diversity of content that users seek out organically in other sections of app. These results suggest that personalized recommendation algorithms have the potential to affect users' preferences, and may play a role in Balkanizing online content consumption.

While \citet{lee2019recommender} find that recommender systems have a neutral-to-positive effect on individual-level diversity and decrease aggregate diversity, we find the opposite: in our setting, personalized recommendations \textit{decreased} individual-level diversity and \textit{increased} aggregate diversity. We believe there are multiple reasons this may be the case. First, as argued by \citet{lee2019recommender}, the effect of recommender systems is likely dependent on both the particular algorithm used and the setting in which it is deployed. Second, previous economics and management research (e.g., \citet{brynjolfsson2011goodbye}, \citet{lee2019recommender}) has typically measured changes in sales diversity, whereas we measure changes in the distribution of content categories consumed.\footnote{In Appendix \ref{sec:sales_diversity}, we report the effect of the experiment on the ``sales diversity" for podcast consumption.} Given that recommender systems have become a common feature of content platforms, we believe it is important to measure the impact of recommender systems not just on market concentration, but also on the \textit{types} of items that users engage with. Overall, the contrast between previous findings and ours underscores the need to study the effects of many recommendation algorithms, in many contexts, using many different measures of diversity.

Our results suggest multiple interesting extensions. First, while our experiment enables us to measure the effect of short-term exposure to personalized recommendations, we are unable to measure the impact of long-term exposure to personalized recommendations. Long-term exposure may affect content consumption and diversity differently. Second, while category tags provide a coarse sense of the type of content users are consuming, there are other important ways to quantify product diversity. For instance, it may be helpful to measure category similarity, the political skew of a piece of content, or the ``extremity" of a piece of content. Third, it would be worthwhile to more explicitly consider the optimal strategy of a content producer in the presence of recommender systems that affect consumption diversity.
%that pushes users into patterns of consumption that are increasingly individually homogenous and Balkanized.
Finally, in this paper we study personalized recommendations that are solely optimized for engagement. This single objective approach to personalization is common in practice, and our findings suggest that researchers should continue to develop personalization techniques that explicitly take into account the diversity of content recommended to users \citep{marler2004survey, castells2015novelty, lacerda2017multi}.

\section{Conclusion} \label{sec:conclusion}

In this paper, we analyze data from a randomized field experiment and measure the effect of personalized content recommendations not just on the \textit{amount} of content that people consumed, but also on the \textit{diversity} of content that people consumed. We find evidence that an ``engagement-diversity tradeoff" can exist for firms when recommendations are optimized solely to drive engagement. While more personalized recommendations increased user engagement, they also decreased the diversity of content that individual users consume, while simultaneously \textit{increasing} the degree of dissimilarity across users. These shifts in content consumption patterns can negatively impact the rate of churn and average lifetime value for users, and also impact the optimal strategy for content creators. We also find evidence that exposure to personalized content recommendations impacted the types of content that users sought out organically. At first glance, our results are at tension with some recent studies of recommender systems, such as \citet{lee2019recommender}. However, we believe this contrast highlights the need for further experimental studies of recommender systems across a multitude of different business settings and algorithm specifications, as well as the need to develop new methods for measuring the effect of recommender systems. Furthermore, we believe our results underscore the need for researchers to continue developing approaches to personalization that optimize jointly for user engagement and consumption diversity.

%, and that the impact of short-term exposure recommender systems on the volume and diversity of content consumed does not persist once users are no longer exposed to personalized recommendations.

\clearpage
\section{Figures}

\clearpage
\section{Tables}

\input{shows_played_model_popularity.tex}

\input{individual_diversity_model_streamers_only_popularity_no_obs.tex}

\input{shows_played_referrer_popularity.tex}

\input{individual_diversity_model_streams_only_referrer_popularity_no_obs.tex}

\clearpage

\bibliographystyle{informs2014} 
\bibliography{coldstart.bib}

\clearpage

%%TC:ignore

\begin{APPENDICES}

\section{Podcast Categorization}
\label{sec:podcast_categories}

In this section, we provided a more detailed description of how podcasts are categorized on Spotify. In our dataset, there are as many as ten podcast category tags associated with any particular podcast. For instance, the podcast ``Trial By Stone: The Dark Crystal Podcast" has only one category associated with it: ``Arts \& Entertainment". On the other hand, the podcast ``World's Best Parents" has four categories associated with it: ``Comedy," ``Educational", ``Kids \& Family," and ``Stories". Category tags are extracted from podcasts' RSS feeds, and are not determined internally at Spotify. Podcast creators can specify as many category tags for a podcast as they wish, although many podcast upload tools limit podcast creators to three categories. Of the category tags podcast creators specify, no one category is identified as a ``primary" category. Podcast creators are incentivized to select truthful category tags for their shows, since inaccurate tags can lead to their shows being removed from important venues, such as the iTunes store. 

Figure \ref{fig:podcast_categories_app} shows the podcast section of the ``Browse" pane on Spotify's desktop app, which allows users to browse through podcasts by selecting one of the thirteen categories. Figure \ref{fig:inset_descriptive_plot_popularity} shows the distribution of categories per podcast for podcasts available on Spotify as of July 8th, 2019, as well as the fraction of all podcasts that have each category tag associated with them.\footnote{Actual podcast category names are removed due to confidentiality concerns.} 68.23\% of podcasts have only one category associated with them, and 97.50\% of podcasts have three or fewer categories associated with them. The most common podcast category is associated with 30.34\% of all podcasts, whereas the least common category is associated with 0.20\% of all podcasts. 

As mentioned in the main text, when a podcast has multiple category tags associated with it, we divide that podcast's streams evenly across all of the categories with which it is associated. For example, if a user streamed an episode of ``World's Best Parents", this would count as 0.25 streams for each of the four categories with which ``World's Best Parents" is associated.
% Per Daniel -- test for robustness to this choice. One way to do this is to do a version of the analysis where each podcast can only have one category, just randomly select from the n tags associated with the podcast.

\section{Bucket Randomization Procedure \& Balance Checks}
\label{sec:randomization}

 In this section, we describe the randomization procedure used to assign users to the treatment and control arms of the experiment we analyze, and report the balance of different user-level covariates across the two treatment conditions.
 
 Spotify users were assigned to treatment arms using the following procedure: every Spotify user is first assigned to one of ten thousand ``buckets" based on a hash of their Spotify username. An equal number of these buckets were randomly assigned to the treatment and control conditions. Each user receives the treatment corresponding to their bucket. A subset of buckets are also labeled as ``long-term hold out" buckets, and are not included in Spotify experiments conducted on the ``Home" section of the app. ``Long-term hold out" buckets assigned to our treatment and control conditions were not shown the ``Podcasts to Try" shelf, and are not included in our analysis. This assignment procedure results in 405,401 treatment users across 86 buckets, and 447,536 control users across 94 buckets. Critically, at the time of the experiment, Spotify did not create new user buckets each time an experiment was launched, which means that users within a given treatment assignment bucket share a treatment assignment history for previous experiments. To account for this, we report either cluster-robust standard errors or cluster bootstrap standard errors for all experiment analyses.

Table \ref{tab:summary_stats} shows bucket-level summary statistics for user buckets in both the treatment and control conditions, and tests for statistically significant differences between them. With the exception of average number of exposed users per bucket, we do not find statistically significant differences between the control group and treatment group for any of the specified user-level covariates. We believe that the smaller number of exposed users per bucket in the treatment group is driven by random errors in the generation of recommendations using the neural network-based model. 

\section{Podcast Follows Analysis} \label{sec:app_follows}

In this section, we repeat our analyses for podcast follows, as opposed to podcast streams. Because our results for podcast follows are extremely similar to those for podcast streams, we elect not to conduct referrer-level analysis for podcast follows.

\subsection{Effect on podcast follows}

Figure \ref{fig:inset_shows_follows_plot_popularity} shows the distribution of podcast follows per user during the experiment in both treatment arms, and Table \ref{tab:shows_followed_model} reports the estimated effect of the treatment on podcast follows per user during the duration of the experiment, both with and without controlling for user-level covariates. We find that the treatment increased the number of podcast follows per user by 51.38\% ($\pm$ 7.64\%). 

Using the principal stratification approach \citep{frangakisPrincipalStratificationCausal2002, ding2017principal}, we are able to measure the extent to which this treatment effect is driven by compositional shifts, as opposed to intensity shifts. We find that the treatment led ``compliers" to follow 1.499 (95\% CI: (1.472, 1.536)) more podcasts, whereas the treatment led ``always takers" to follow 0.018 (95\% CI: (-0.070, 0.035)) fewer podcasts. In other words, the increase in podcast following in the treatment is driven by a greater number of users following \textit{at least one} podcast during the experiment, as opposed to an increase in the number of podcast follows from those who would have followed a podcast even if they had not been exposed to the treatment.

Table \ref{tab:followed_show_model} reports the estimated effect of the treatment on following at least one podcast during the experiment, both with and without controlling for user-level covariates. We find that the treatment increased the number of Spotify users following at least one podcast by 53.45\% ($\pm$ 5.23\%). 

\subsection{Effect on diversity of podcast follows}

We also measure the effect of the treatment on the individual-level diversity and intragroup diversity for podcast follows.

\subsubsection{Individual-level diversity}

Figure \ref{fig:inset_shows_streamed_diversity_plot_follows} shows the histogram of the user-level Shannon entropy for podcast follows in both treatment arms, and Table \ref{tab:individual_diversity_model_followers_only} reports the difference in the average following user's Shannon entropy, both with and without controlling for user-level covariates. We find that the average Shannon entropy of podcast follows among users who followed at least one podcast was 10.68\% ($\pm$ 2.33\%) lower in the treatment.

However, as was the case for our analysis of streaming behavior, this estimate is non-causal, since we condition on a post-treatment variable (the decision to follow at least one podcast). Using the principal stratification approach \citep{frangakisPrincipalStratificationCausal2002, ding2017principal}, we can identify the causal effect of the treatment on the individual-level diversity of podcast follows for the subset of users who would follow a podcast, regardless of which treatment condition they were exposed to (i.e., ``always takers"). We estimate that on average, the treatment decreased the individual-level diversity of always takers by 0.067 (95\% CI: (0.052, 0.081)). In other words, the causal effect of the treatment on the individual-level diversity of podcast follows was negative for ``always takers."

Table \ref{tab:individual_diversity_model_all_follows} reports the estimated effect of the treatment on the individual-level diversity of podcast follows for all users in the experiment, both with and without controlling for user-level covariates. We find that the treatment increased the Shannon entropy for podcast follows by 37.06\% ($\pm$ 6.17\%). Figure \ref{fig:user_level_diversity_histogram_follows_only_by_follows} shows histograms of the user-level Shannon entropy for podcast follows in both treatment arms conditional on a user following a particular number of podcasts during the experiment.

% Principal strat CACE point estimate: 0.578
% Principal strat CACE 95% CI: (.566, .591)

\subsubsection{Effect on intragroup diversity}

We find that the treatment increased the intragroup diversity for follows by 7.12\% (95\% CI: (6.04\%, 8.23\%), from 0.687 (95\% CI: (0.681, 0.693)) in the control group to 0.736 (95\% CI: (0.732, 0.740))

%The difference in intragroup diversity between the treatment and control is 0.049 (95\% CI: (0.042, 0.056))

\subsection{Long-term effects}

We use data collected between May 3, 2019 and July 17, 2019 to test for longer-term effects of the treatment. We repeat our main analyses on cross-sectional datasets that describe users' behavior over time intervals spanning from the 3rd of the month to the 17th of the month, and from the 18th of the month to the 2nd of the month.

Figure \ref{fig:compare_time_series_follows_plot_popularity} shows the long-term effect of the treatment on the average number of podcast follows per user, the average Shannon entropy of podcast follows conditional on following at least one podcast, and the intragroup diversity of podcast follows. Across all of these outcomes, we observe the same trend: the large treatment effects observed during the experiment quickly shrink in magnitude, and in some cases disappear entirely, once the experiment has concluded. Figure \ref{fig:users_following_dynamic} shows the long-term effect of the treatment on the number of users following at least one podcast. Figure \ref{fig:individual_diversity_follows} shows the long-term effect of the treatment on the individual-level diversity for podcast follows across all users in our sample. For both of these time series, we also observe the rapid dissipation of treatment effects.

\subsection{Effect on ``sales diversity"}

Figure \ref{fig:inset_gini_plot_follows_popularity} shows the Lorenz curve for podcast follows across the top 200 podcasts in both treatment arms of the experiment. The difference between the two curves indicates that the treatment makes podcast following \textit{less} concentrated, and distributes a larger fraction of follows to less popular podcasts, i.e., the treatment increases the sales diversity for podcast follows. We confirm this by measuring the Gini coefficients corresponding to each treatment arm's Lorenz curve. We find that that the treatment reduces the Gini coefficient by 0.138 (95\% CI: (0.116, 0.149)), from 0.588 to 0.450.

We also measure the effect of the treatment on sales diversity by estimating Equation \ref{eq:streams_equation} with follow counts and follow rank, as opposed to stream counts and stream rank. Figure \ref{fig:ln_ln_plot_immediate_follows} shows the relationship between $\ln(Follows_i + 1)$ and $\ln(Follows\,Rank_i)$ across all podcasts appearing in our dataset, and Table \ref{tab:ln_rank_model_follows} shows the results of estimating Equation \ref{eq:streams_equation} on data from the top 200 podcasts in each treatment arm. The reported 95\% confidence intervals are calculated with the cluster bootstrap ($n_{boot}$ = 1,000). The positive estimate for $\beta_3$ also indicates that the treatment \textit{increases} sales diversity. 

\section{Long-term Treatment Effects}
\label{sec:long_term_effects}

%, and also differences in the effect of the treatment over time. 

%Furthermore, all of the large treatment effects that we observe dissipate quickly once the experiment has ended, indicating both that users tend to revert to their original listening habits once personalized recommendations are no longer shown, and that firms must constantly recommend new content to users in order to achieve a long-term increase in user engagement.

%, and on the long-run impact of short-term exposure to personalized recommendations. 

%Furthermore, all of the treatment effects that we observe shrink in magnitude or disappear entirely once users are no longer exposed to personalized recommendations. Taken together, 

%, users revert to their baseline counterfactual level of content diversity once no longer exposed. 

%, and that the impact of short-term exposure recommender systems on the volume and diversity of content consumed does not persist once users are no longer exposed to personalized recommendations.

In this subsection, we use data collected between May 3, 2019 and July 17, 2019 to test for longer-term effects of the treatment. We repeat our main analyses on cross-sectional datasets that describe users' behavior over time intervals spanning from 3rd of the month to the 17th of the month, and from the 18th of the month to the 2nd of the month. Testing for long-term effects allows us to determine whether short-term exposure to personalized podcast recommendations has a lasting impact on the types of content that users consume, or if users revert to their their counterfactual baseline podcast listening once individually personalized recommendations are no longer shown.

Figure \ref{fig:compare_time_series_streams_plot_popularity} shows the long-term effect of the treatment on the average number of podcast streams per user, the average Shannon entropy of podcast streams conditional on streaming at least one podcast, and the intragroup diversity of podcast streams. Across all of these outcomes, we observe the same trend: the large treatment effects observed during the experiment quickly shrink in magnitude, and in some cases disappear entirely, once the experiment has concluded.

We also measure the long-term effect of the treatment on podcast streams originating from different referral surfaces. This allows us to identify potential heterogeneity in the extent to which short-term exposure to personalized podcast recommendations has a long-term effect on consumption habits across both recommended listening and organic listening. Figure \ref{fig:compare_time_series_streams_referrer_plot_popularity} shows the long-term effect of the treatment on average podcast streams per user, Shannon entropy for streams conditional on streaming at least one podcast, and intragroup diversity for streams originating from both home and non-home surfaces. Stream referrer-level treatment effects follow the same trend as overall effects, and this trend does not vary by stream referrer; treatment effects dissipate quickly, or disappear entirely, once the experiment has ended. The lack of long-term treatment effects suggests that short-term exposure to personalized podcast recommendations does not affect long-term listening behavior through algorithmic spillovers or changes in what users seek out organically. 

It is worth noting that the number of podcast streams per user over time is dependent on users being exposed to podcast content in the ``Home" section of the Spotify app. However, the number of impressions that podcast content received on ``Home" varied considerably in the months following the experiment. During the experiment, the majority of podcast content impressions on ``Home" came from the ``Podcasts to Try" shelf, since it was anchored in the second slot. After the experiment had ended, the ``Podcasts to Try" shelf was briefly removed from the Spotify app to be productionized. The treatment version of the shelf was relaunched to 100\% of Spotify users in late May, however, the shelf was no longer anchored in the second slot. As a result, there were far fewer impressions for all podcast related shelves, including ``Podcasts to try". An experiment to determine the optimal amount of boosting for podcast shelves was launched in mid-May, and podcast shelf boosting was launched to 100\% of users in early June. Figure \ref{fig:coldstart_shelf_impressions_simple_treatment} shows the number of impressions for podcast content on both ``Podcasts to Try" and other podcast-related shelves for both treatment and control users over time. Note that the time series for the two experiment treatment arms are essentially identical.

\section{Effect of the treatment on distinct podcast streamers}
\label{sec:podcast_streamers}

In this section, we report the effect of the treatment on the number of users streaming at least one podcast during the experiment. 

Table \ref{tab:played_show_model} reports the estimated effect of the treatment, both with and without controlling for user-level covariates. We find that the treatment increased the number of Spotify users streaming podcasts by 36.33\% ($\pm 3.01\%$). Table \ref{tab:played_shows_model_referrer} reports the estimated effect of the treatment on streaming at least one podcast during the experiment from either type of referral surface, both with and without controlling for user-level covariates. We find that the treatment increased the number of Spotify users streaming podcasts from ``Home" by 59.17\% ($\pm 4.58\%$), and the number of users streaming podcasts from non-home surfaces by 12.55\% ($\pm 2.94\%$). 

Figures \ref{fig:users_streaming_dynamic} and \ref{fig:user_level_streamers_surface_plot} show the long-term effect of the treatment on the number of users streaming at least one podcast, both overall and conditional on streaming surface. We find that the large treatment effects we observe during the experiment rapidly dissipate once the experiment has concluded.

\section{Principal stratification methodology}
\label{sec:principal_strat}

In this section, we describe our principal stratification methodology, which is based on the principal stratification approach described by \citet{frangakisPrincipalStratificationCausal2002} and \citet{ding2017principal}. 

The principal stratification framework allows for causal inference in cases where an intermediate variable (in our case, listening to or following at least one podcast) leads to sample selection issues. Using this framework, we are able to separately measuring causal effects of the treatment for ``always takers," i.e., those would stream or follow a podcast, regardless of their treatment status and ``compliers," i.e., those who would follow or stream a podcast only if treated. The key assumption necessary for implementing principal stratification is weak general principal ignorability \citep{ding2017principal}, which states that the expected outcome conditional on the intermediate variable (streaming or following at least one podcast) is independent of strata (complier, always taker, never taker) after controlling for covariates. 

Our implementation of the principal stratification framework uses the marginal method described by \citet{feller2017principal} to compute the probability that each user in our sample is a complier, always taker, or never taker. Under the principal stratification approach's monotonicity assumption, we can assume that users who do not stream or follow a podcast in the treatment are never takers, and that podcast streamers or followers in the control are always takers. For all other users, we estimate the probability that they are an always taker using a logistic regression model that is trained on control data and predicts streaming or following a podcast using user-level covariates. Similarly, we estimate the probability that a user is a never taker using a logistic regression model that is trained on treatment data and predicts streaming or following a podcast using user-level covariates. Once we have estimated $P(always\,taker)_i$ and $P(never\,taker)_i$, we can calculate $P(complier)_i$, since $P(complier)_i = 1 - P(always\,taker)_i - P(never\,taker)_i$. In cases where $P(always\,taker)_i + P(never\,taker)_i > 1$, we set $P(complier)_i = 0$ and normalize the other two probabilities so that they sum to 1. In both logistic regression models, user age bucket, user gender, and user account age (in days) are the covariates used to predict the intermediate variable.\footnote{We also calculate strata probability estimates using the EM algorithm described by \citet{ding2017principal}. The point estimates obtained using this method are qualitatively similar to those obtained using the marginal method. However, we choose the marginal method for computational tractability when calculating bootstrap standard errors.} Once we have computed the probability that each user belongs to each stratum, we use these probabilities as weights to construct causal stratum-level treatment effect estimators. Confidence intervals are calculated using a clustered bootstrap ($n_{boot} = 1,000$).

We test that the principal stratification model we have proposed is accurate using the balancing conditions proposed by \citet{ding2017principal}. Simply put, the balancing conditions require that within each stratum, the treatment should not appear to have a causal effect on any function of the pretreatment covariates used to estimate a given unit's stratum. For both intermediate outcomes (streaming at least one podcast and following at least one podcast), we estimate the effect of the treatment on each pre-treatment user-level covariate in each stratum. The results for podcast streaming are shown in Figure \ref{fig:balance_checks_streams_plot_diversity_popularity} and the results for podcast following are shown in Figure \ref{fig:balance_checks_follows_plot_diversity_popularity}. In both cases, the estimated effects are nearly zero across all strata and covariates, indicating that the balancing conditions are satisfied. 

\section{Effect of the treatment on individual-level diversity (all users)}
\label{sec:ind_div_all}

In this section, we report the effect of the treatment on the individual-level diversity when including all users in our analysis, regardless of whether or not they streamed any podcasts during the experiment. Table \ref{tab:individual_diversity_model_all} reports the estimated effect of the treatment on the Shannon entropy, both with and without controlling for user-level covariates. We find that the treatment increased the average Shannon entropy for podcast streams by 21.16\% ($\pm$ 2.89\%). Table \ref{tab:individual_diversity_model_all_streams_referrer} reports the estimated effect of the treatment on the referrer-specific user-level Shannon entropy, both with and without controlling for user-level covariates. We find that the treatment increased the average Shannon entropy of home streams by 29.83\% ($\pm$ 3.76\%), and increased the average Shannon entropy of non-home streams by 7.36\% ($\pm$ 3.19\%). 

Figures \ref{fig:individual_diversity_streams} and \ref{fig:diversity_streamers_all_surface} show the long-term effect of the treatment on the individual-level diversity of podcast streams across all users in the experiment, both overall and conditional on streaming surface. We find that the large treatment effects we observe during the experiment rapidly dissipate once the experiment has concluded.

\section{Effect on ``sales diversity"}
\label{sec:sales_diversity}

In this section, we measure the effect of the treatment on the ``sales diversity" of podcast consumption, as measured through the Lorenz curve and Gini coefficients corresponding to podcast streaming in both treatment arms of the experiment.

Figure \ref{fig:lorenz_streams} shows the Lorenz curve for podcast streaming across the top 1,000 podcasts in both treatment arms of the experiment. The difference between the two curves indicates that the treatment makes podcast streaming \textit{less} concentrated, and distributes a larger fraction of streams to less popular podcasts. In other words, the treatment increases sales diversity. We confirm this by measuring the Gini coefficients corresponding to each treatment arm's Lorenz curve. We find that that the treatment reduces the Gini coefficient by 0.050 (95\% CI: 0.037, 0.061), from 0.692 to 0.642.

We also measure the effect of the treatment on sales diversity by estimating the following model \citep{brynjolfsson2011goodbye}:

\begin{equation} \label{eq:streams_equation}
\begin{split}
\ln(Streams_i + 1)  = & \beta_0 + \beta_1 \ln(Streams\,Rank_i) + \beta_2 Treatment_i + \\
& \beta_3 Treatment_i \times \ln(Streams\,Rank_i) + \epsilon_i,
\end{split}
\end{equation}

\noindent where $Streams_i$ is how many streams podcast $i$ received during the experiment in a particular treatment arm, $Streams\,Rank_i$ is podcast $i$'s rank among all podcasts in that treatment arm, and $Treatment_i$ indicates the treatment arm corresponding to the observation. The coefficient of interest is $\beta_3$, which tests for a difference between the two treatment arms in the rate at which number of streams decreases with stream rank.

Figure \ref{fig:ln_ln_plot_immedate} shows the relationship between $\ln(Streams_i + 1)$ and $\ln(Streams\,Rank_i)$ across all podcasts appearing in our dataset, and Table \ref{tab:ln_rank_model} shows the results of estimating Equation \ref{eq:streams_equation} on data from the top 1,000 podcasts in each treatment arm. The reported 95\% confidence intervals are calculated with the cluster bootstrap ($n_{boot}$ = 1,000). The positive estimate for $\beta_3$ also indicates that the treatment \textit{increases} sales diversity. 

\clearpage
\renewcommand\thefigure{\thesection.\arabic{figure}}    
\section{Additional Figures}
\setcounter{figure}{0}

\begin{figure}[!htbp]
\begin{center} 
\includegraphics[scale=.9]{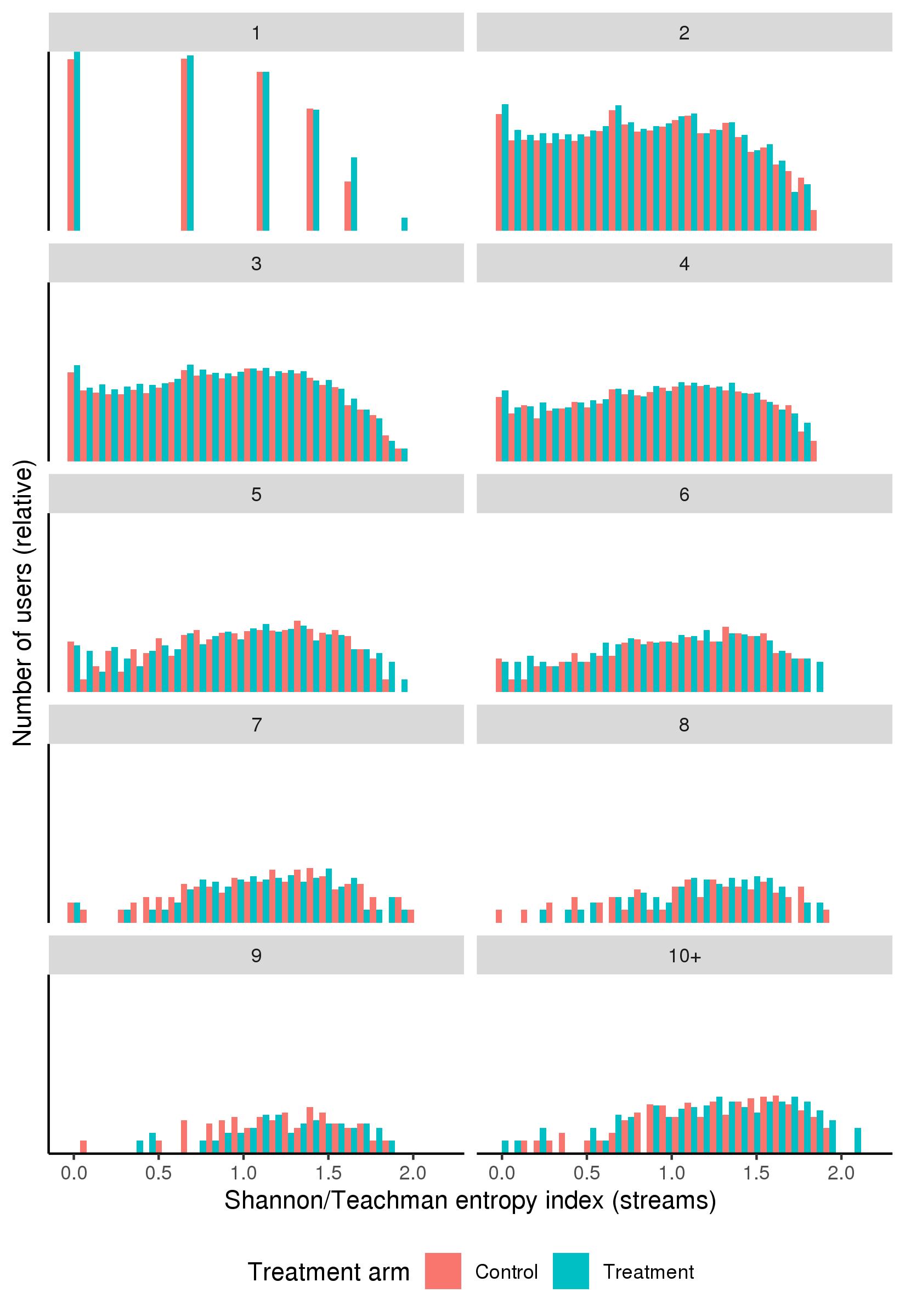}
\caption{The distribution of the user-level diversity for streams in both treatment arms conditional on streaming a set number of podcasts during the experiment. y-axis values are on a log scale, and are hidden due to confidentiality concerns.}
\label{fig:user_level_diversity_histogram_streamers_only_by_streams}
\end{center}
\end{figure}

\begin{figure}[!htbp]
\begin{center}
\includegraphics[height=4in]{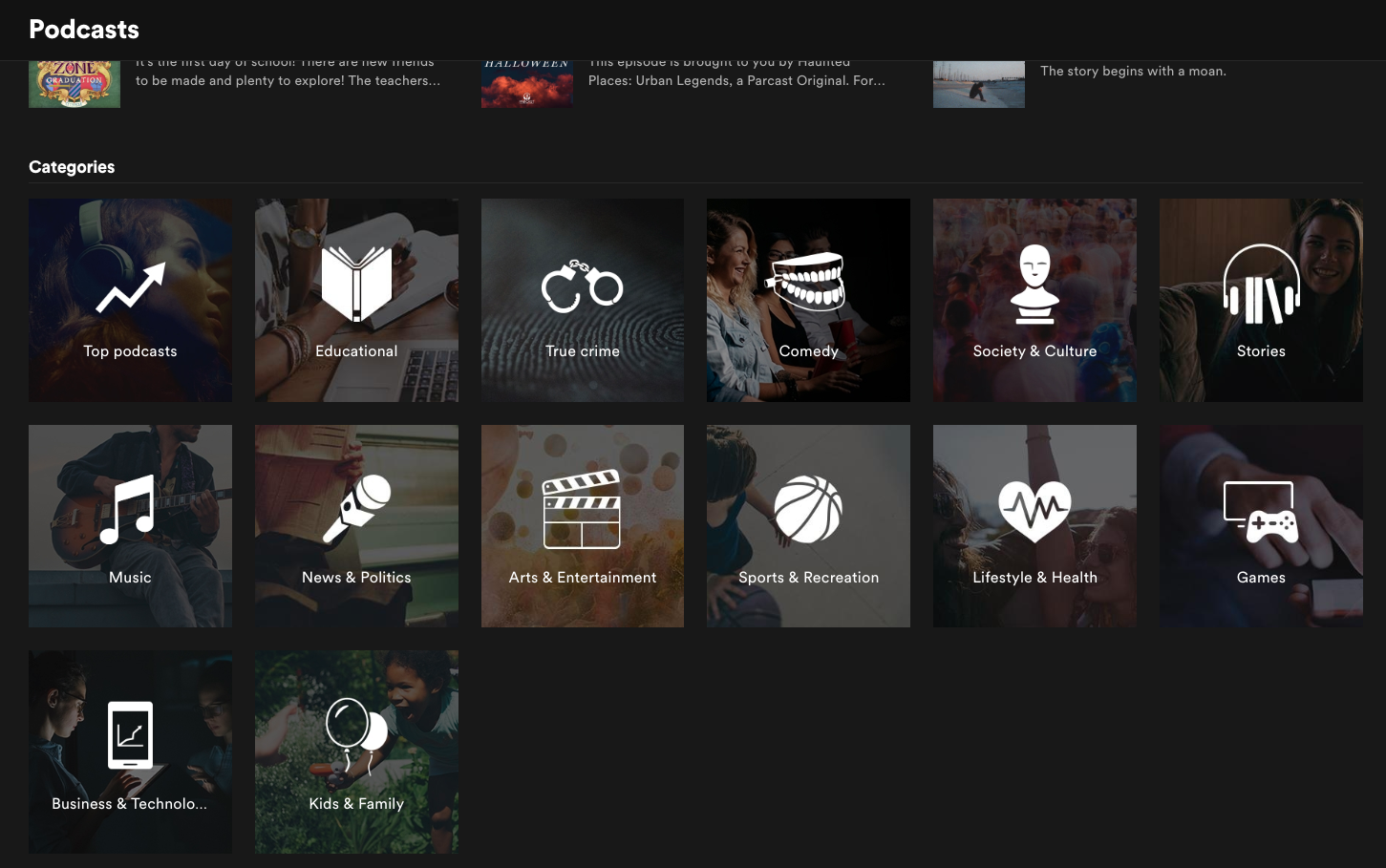}
\caption{Podcast categories on Spotify, as displayed in the Podcasts section of the desktop app's ``Browse" pane.}
\label{fig:podcast_categories_app}
\end{center}
\end{figure}

\begin{figure}[!htbp]
\begin{center}
\includegraphics{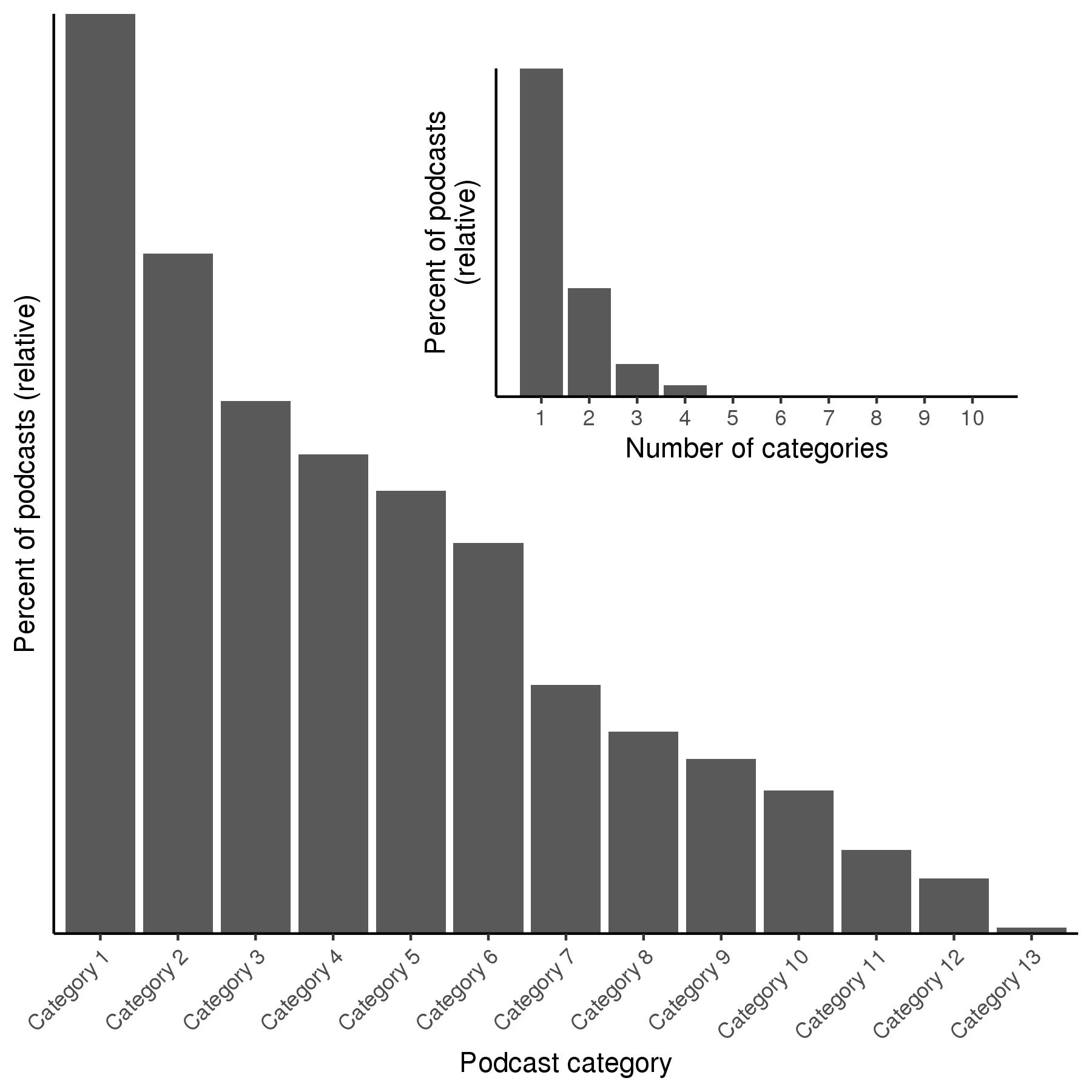}
\caption{Histograms showing the frequency with which each podcast category is attached to a podcast, and the distribution of category tags per podcast. y-axis values and category names hidden due to confidentiality concerns.}
\label{fig:inset_descriptive_plot_popularity}
\end{center}
\end{figure}

\begin{figure}[!htbp]
\begin{center}
\includegraphics{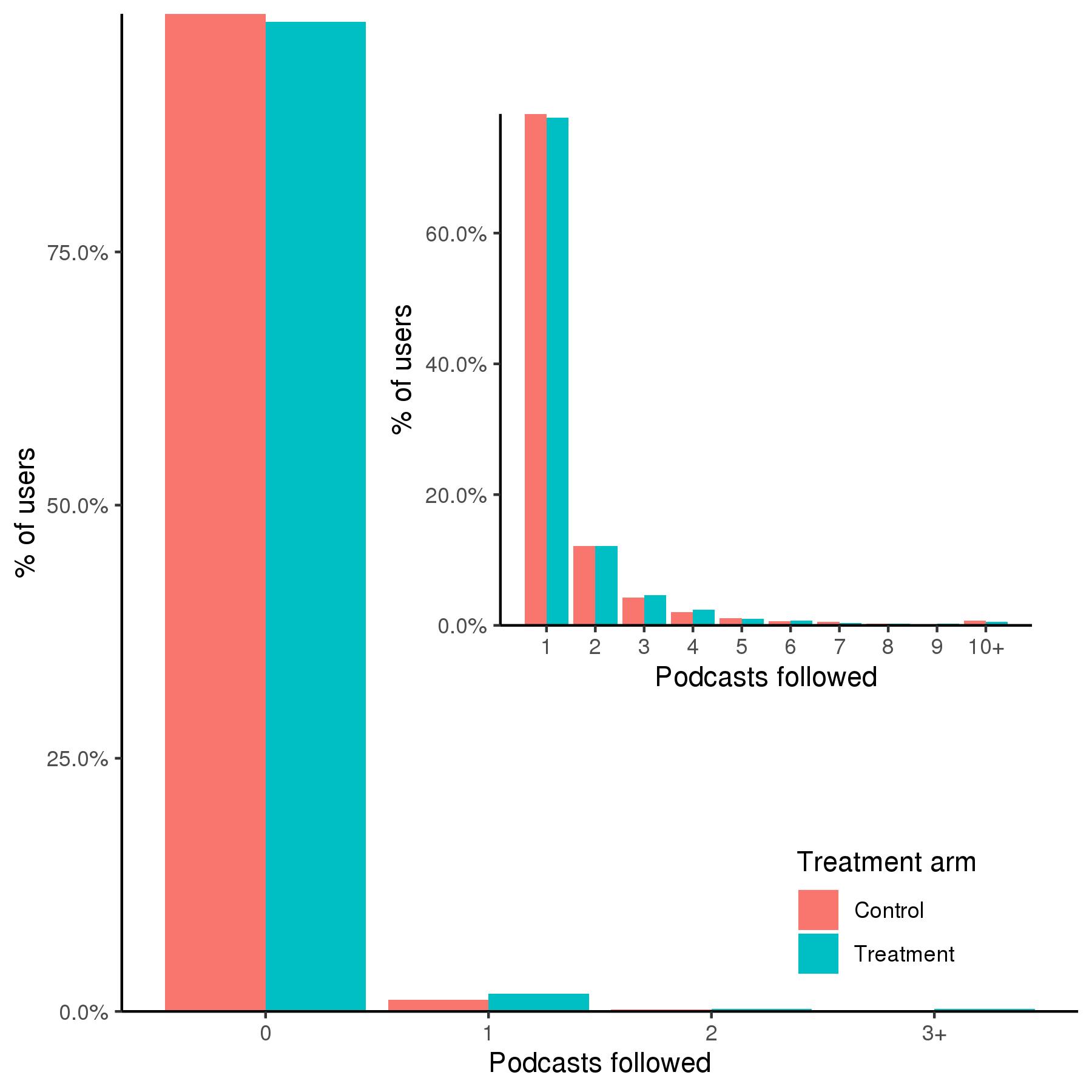}
\caption{The distribution of podcasts followed in both treatment arms. Inset plot shows the distribution of podcasts followed in both treatment arms conditional on following at least one podcast.}
\label{fig:inset_shows_follows_plot_popularity}
\end{center}
\end{figure}

\begin{figure}[!htbp]
\begin{center} 
\includegraphics{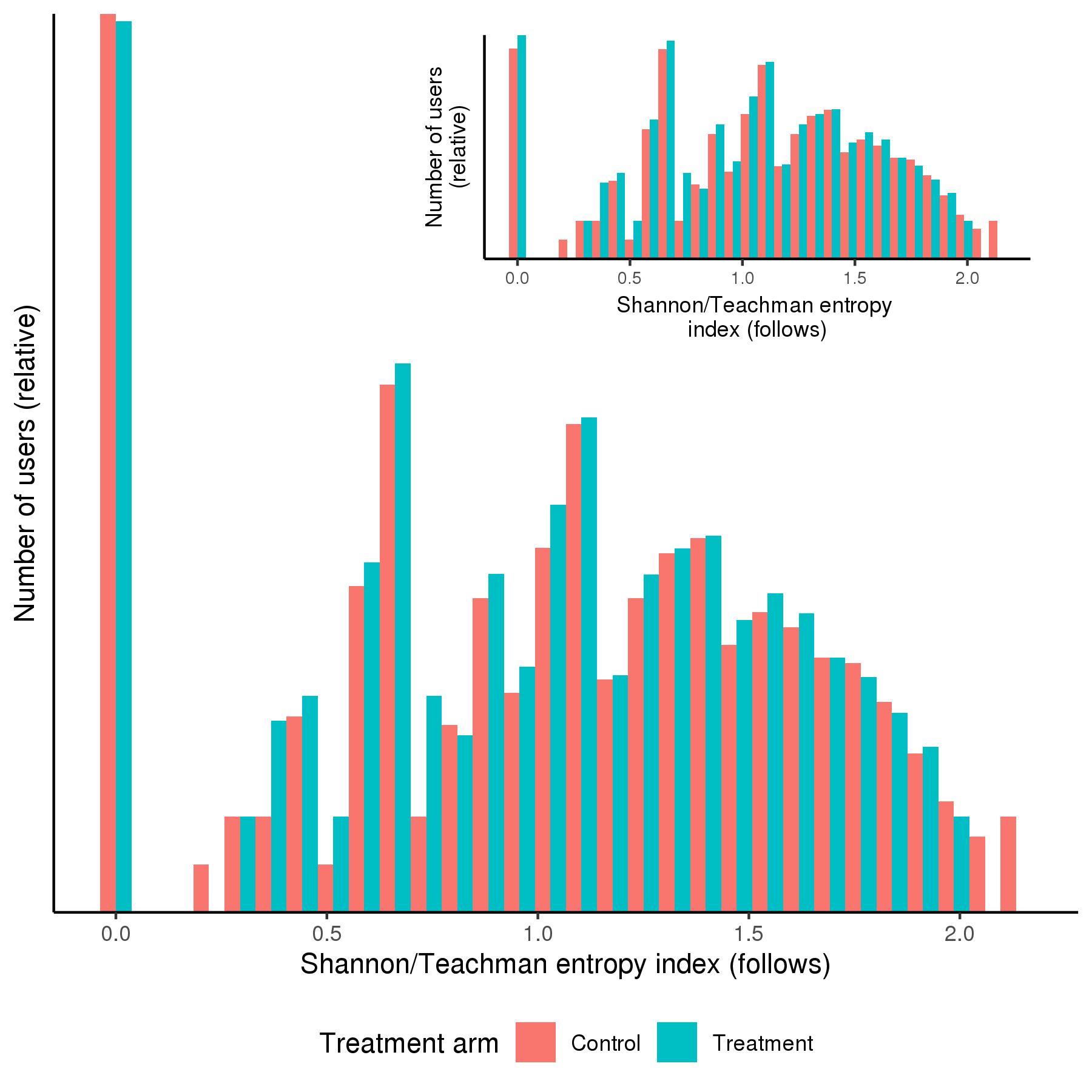}
\caption{The distribution of the user-level diversity for follows in both treatment arms. Inset plot shows the distribution of user-level diversity in both treatment arms conditional on following at least one podcast. y-axis values are on a log scale, and are hidden due to confidentiality concerns.}
\label{fig:inset_shows_streamed_diversity_plot_follows}
\end{center}
\end{figure}

\begin{figure}[!htbp]
\begin{center} 
\includegraphics{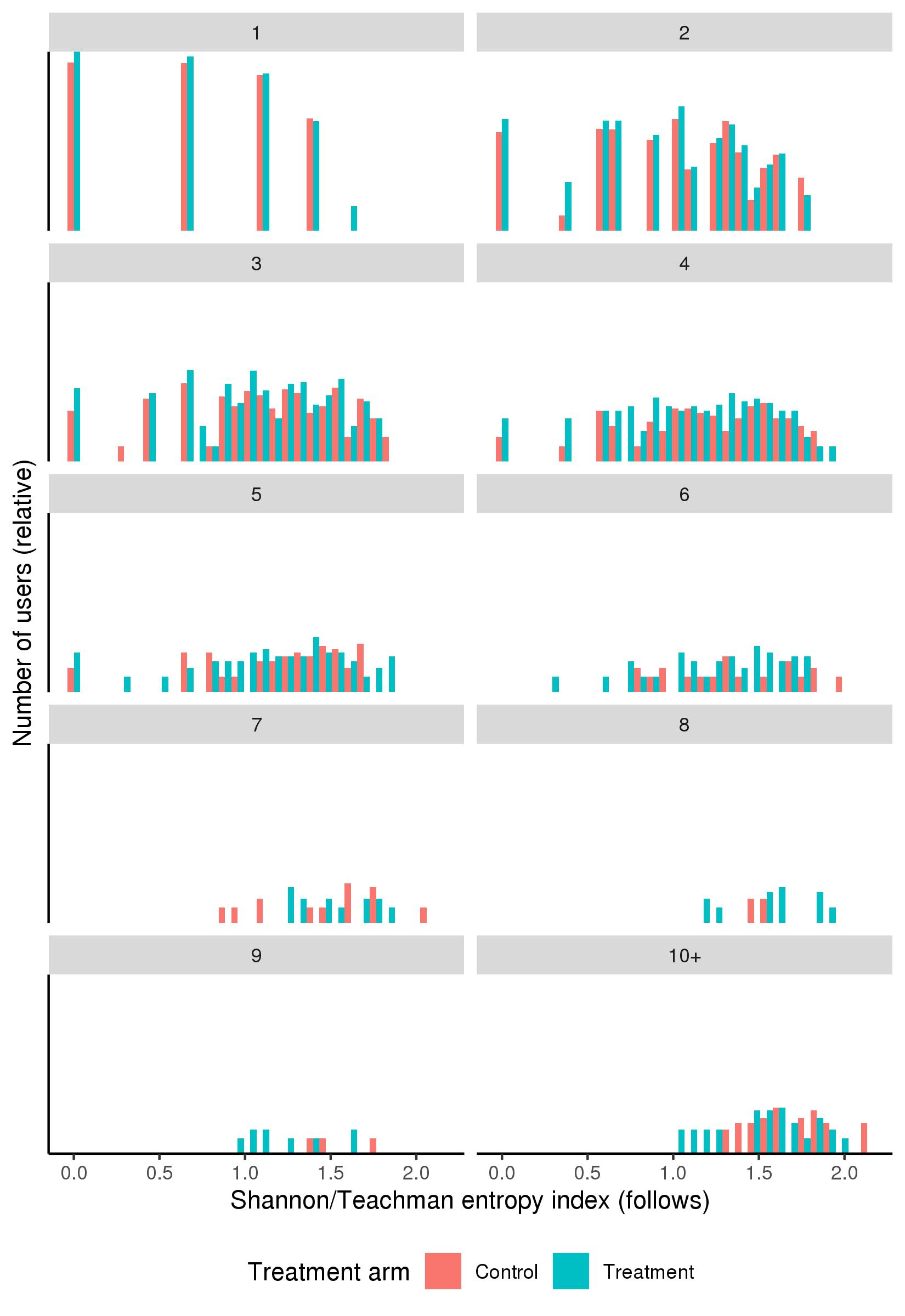}
\caption{The distribution of the user-level diversity for follows in both treatment arms conditional on following a set number of podcasts during the experiment. y-axis values are on a log scale, and are hidden due to confidentiality concerns.}
\label{fig:user_level_diversity_histogram_follows_only_by_follows}
\end{center}
\end{figure}

\begin{figure}[!htbp]
\begin{center} 
    \includegraphics{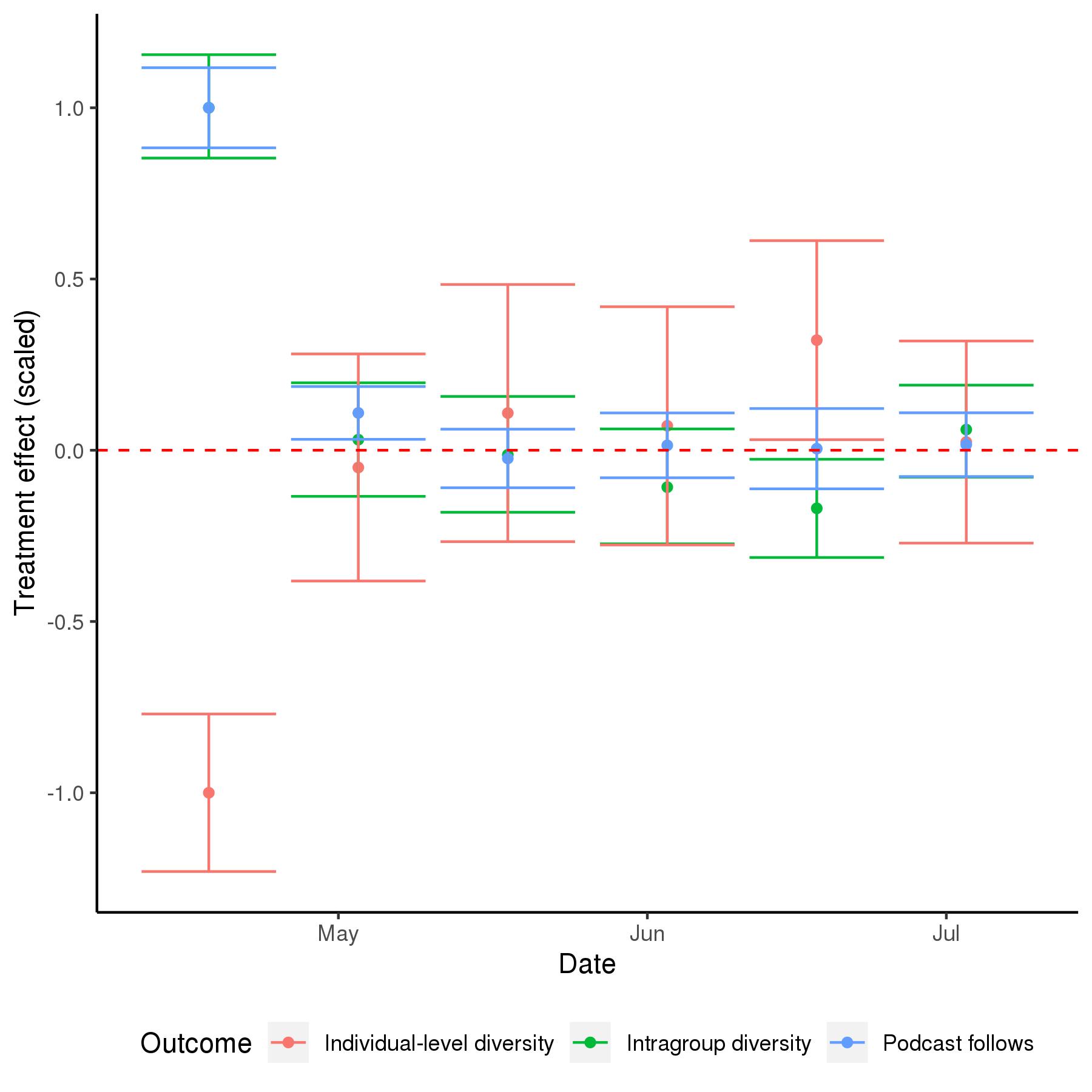}
    \caption{The long-term effect of the treatment on podcast follows per user, individual-level following diversity conditional on following at least one podcast, and intragroup following diversity. Each outcome's time series is scaled by the absolute value of the magnitude of the treatment effect during the experiment.}
    \label{fig:compare_time_series_follows_plot_popularity}
\end{center}
\end{figure}

\begin{figure}[!htbp]
\begin{center} 
\includegraphics{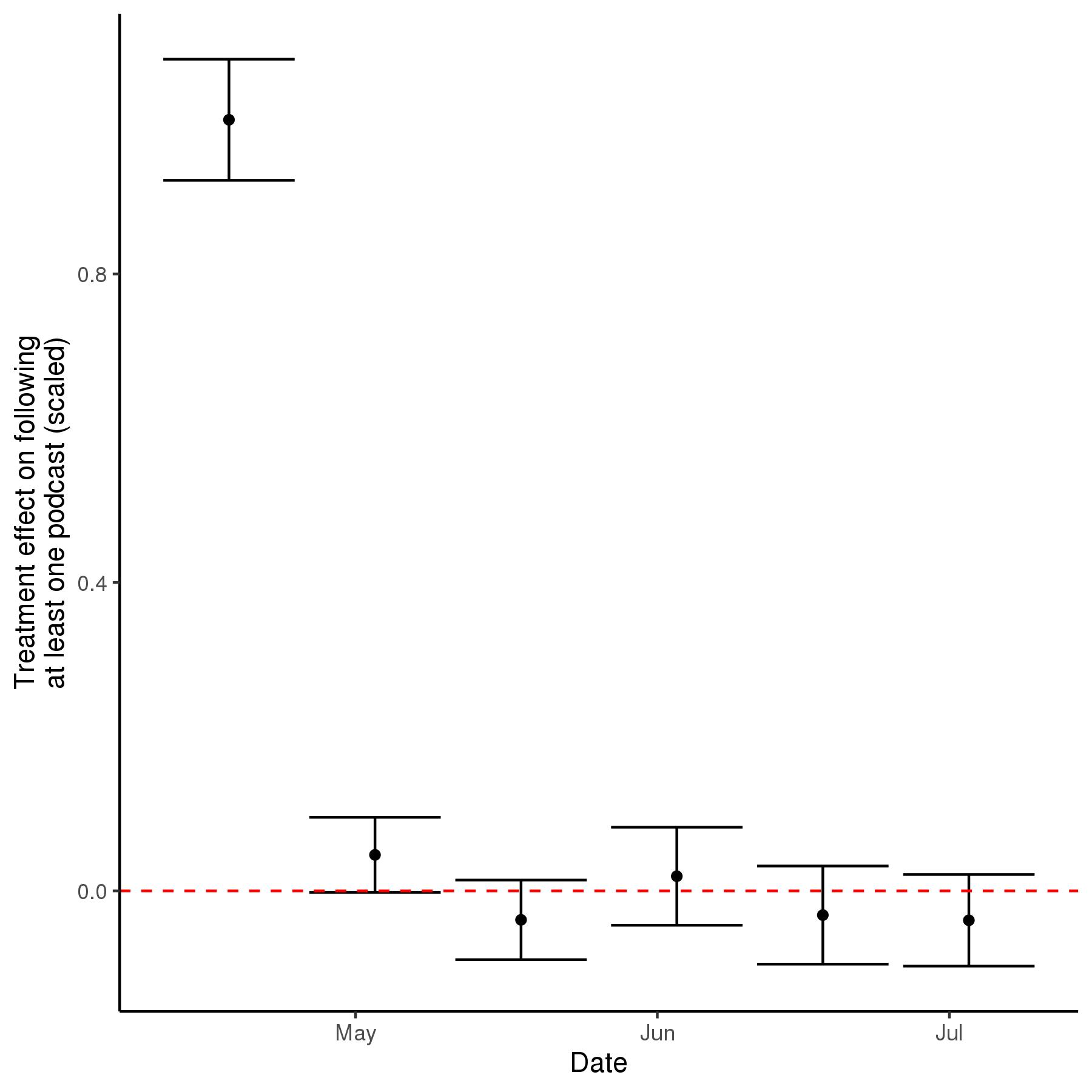}
\caption{The long-term effect of the treatment on the percentage of users following at least one podcast. The time series is scaled by the absolute value of the magnitude of the treatment effect during the experiment.}
\label{fig:users_following_dynamic}
\end{center}
\end{figure}

\begin{figure}[!htbp]
\begin{center} 
    \includegraphics{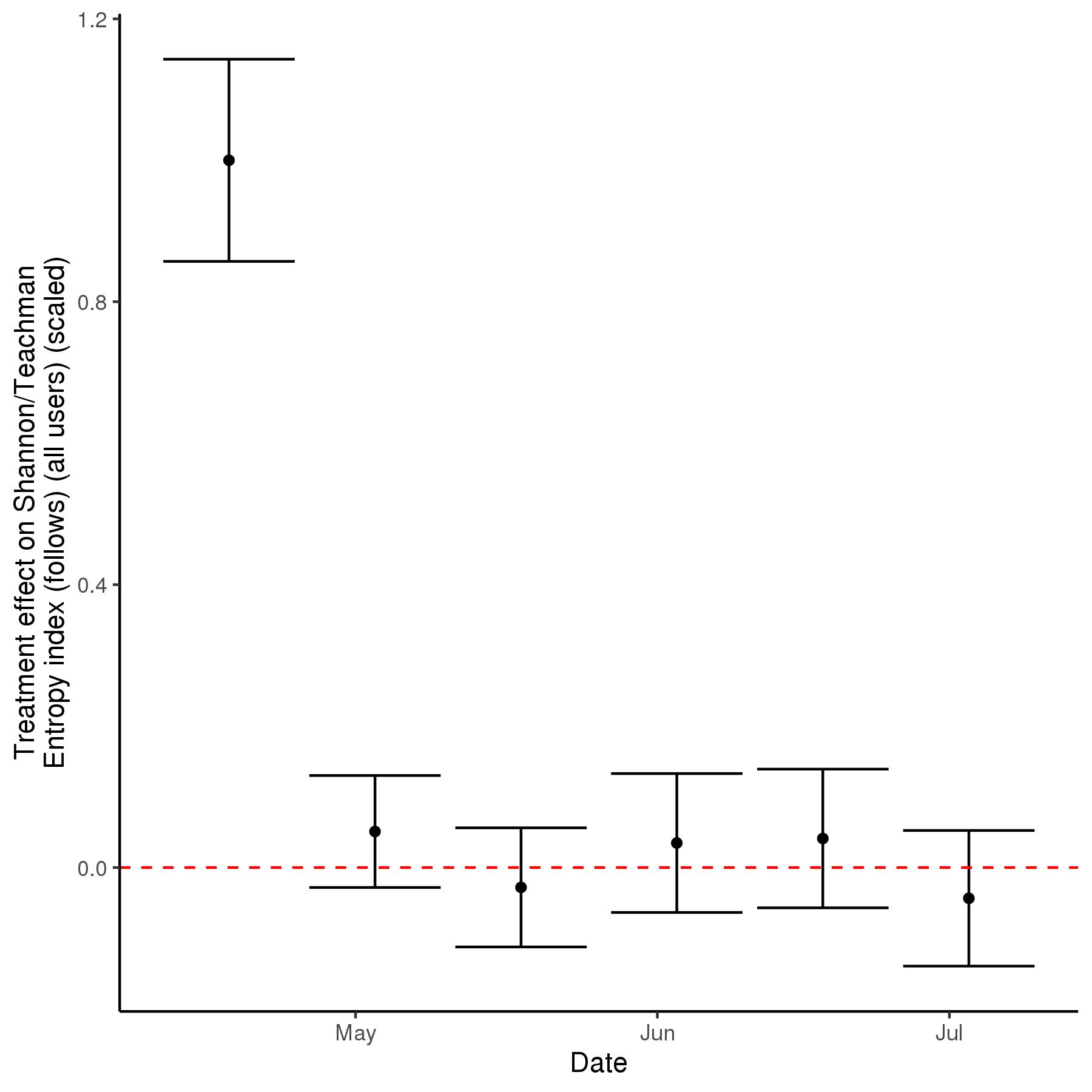}
    \caption{The long-term effect of the treatment on the average user-level Shannon entropy for follows. The time series is scaled by the absolute value of the magnitude of the treatment effect during the experiment.}
    \label{fig:individual_diversity_follows}
\end{center}
\end{figure}

\begin{figure}[!htbp]
\begin{center} 
\includegraphics{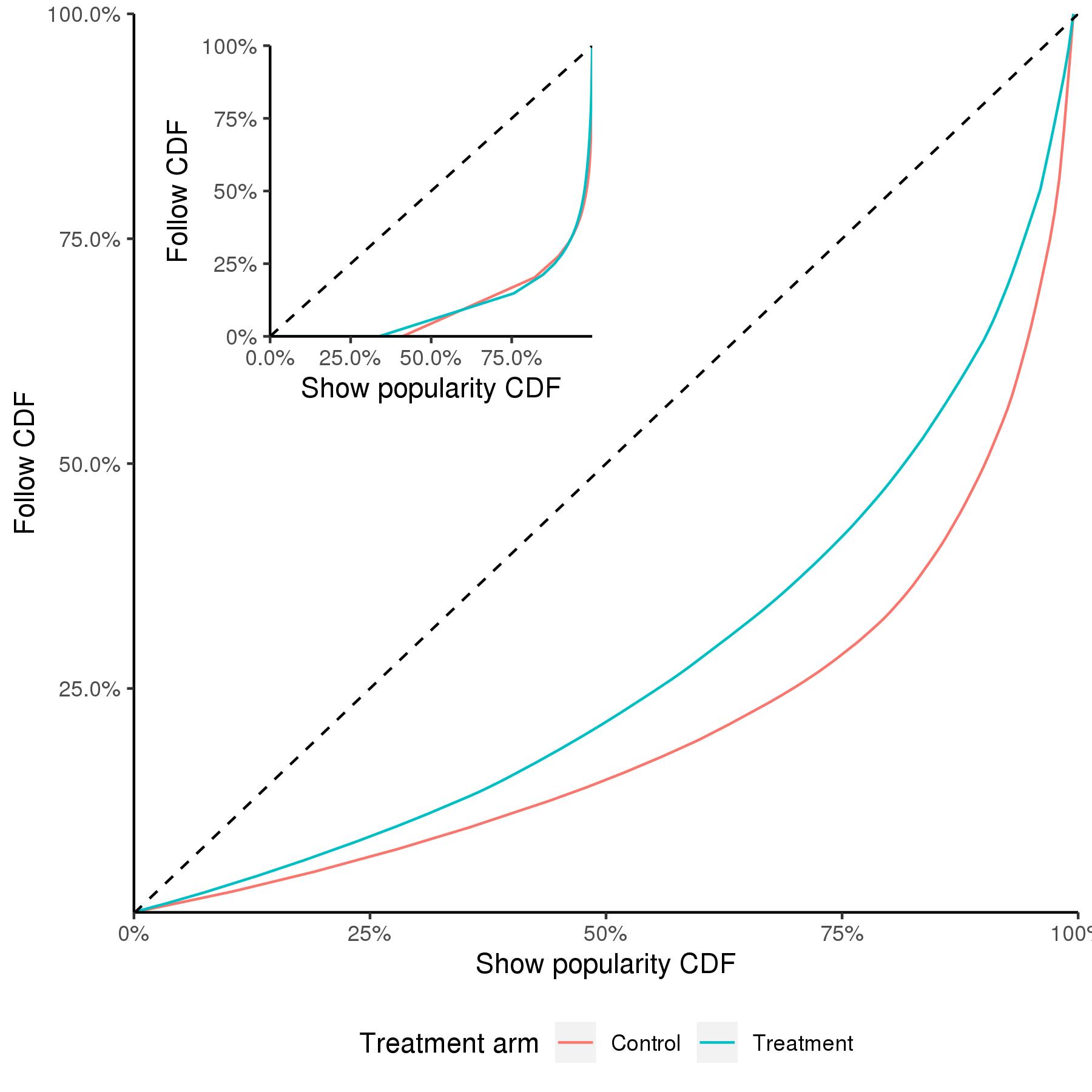}
\caption{The Lorenz curves for podcast follows, calculated separately for users in the treatment and control. The data for each Lorenz curve is limited to the 200 most followed podcasts in the corresponding treatment arm data. The inset curve shows the Lorenz curve for follows across all podcasts.}
\label{fig:inset_gini_plot_follows_popularity}
\end{center}
\end{figure}

\begin{figure}[!htbp]
\begin{center} 
\includegraphics{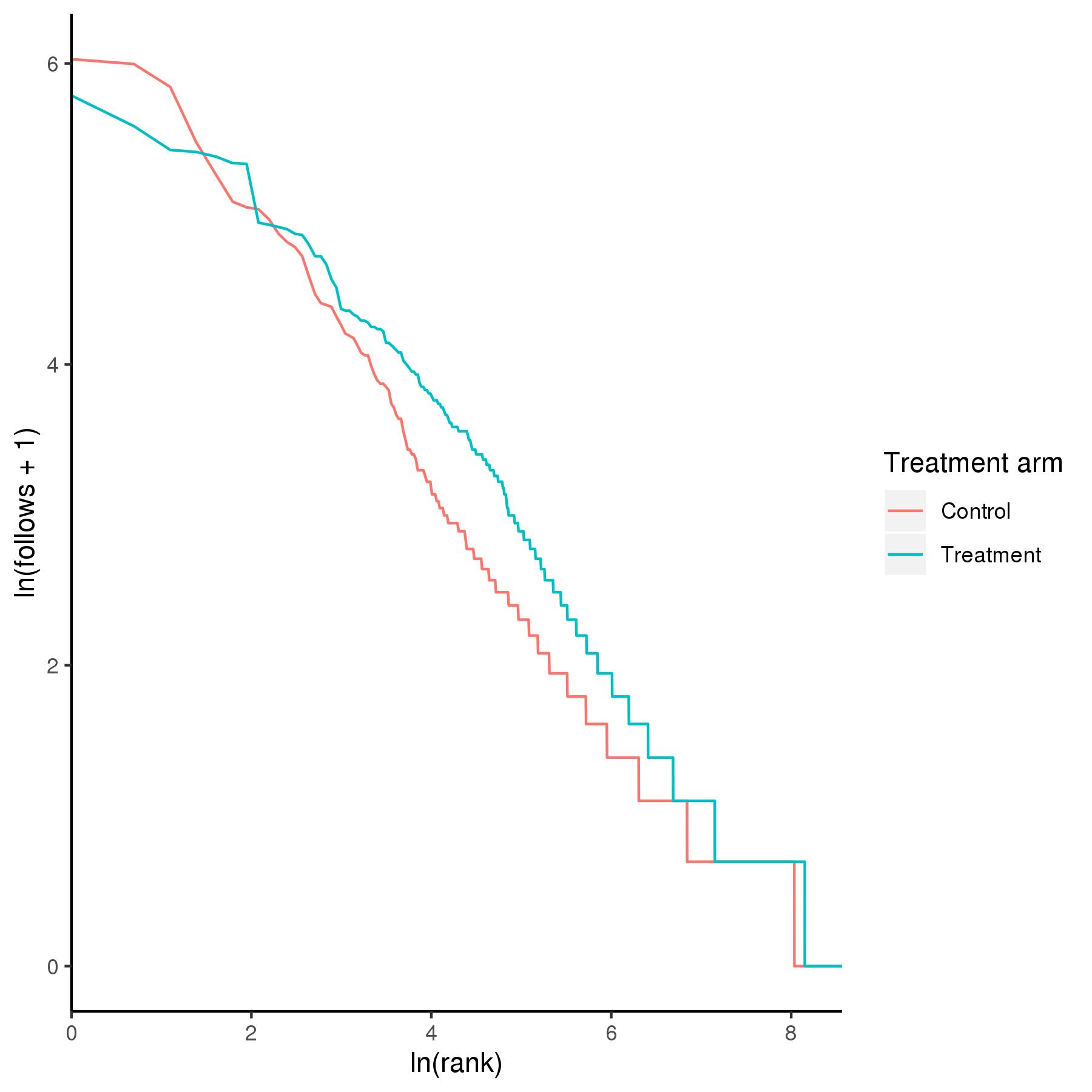}
\caption{The relationship between ln(follows + 1) and ln(follow rank) for both the control and treatment arms of the experiment.}
\label{fig:ln_ln_plot_immediate_follows}
\end{center}
\end{figure}

\begin{figure}[!htbp]
\begin{center}
\includegraphics{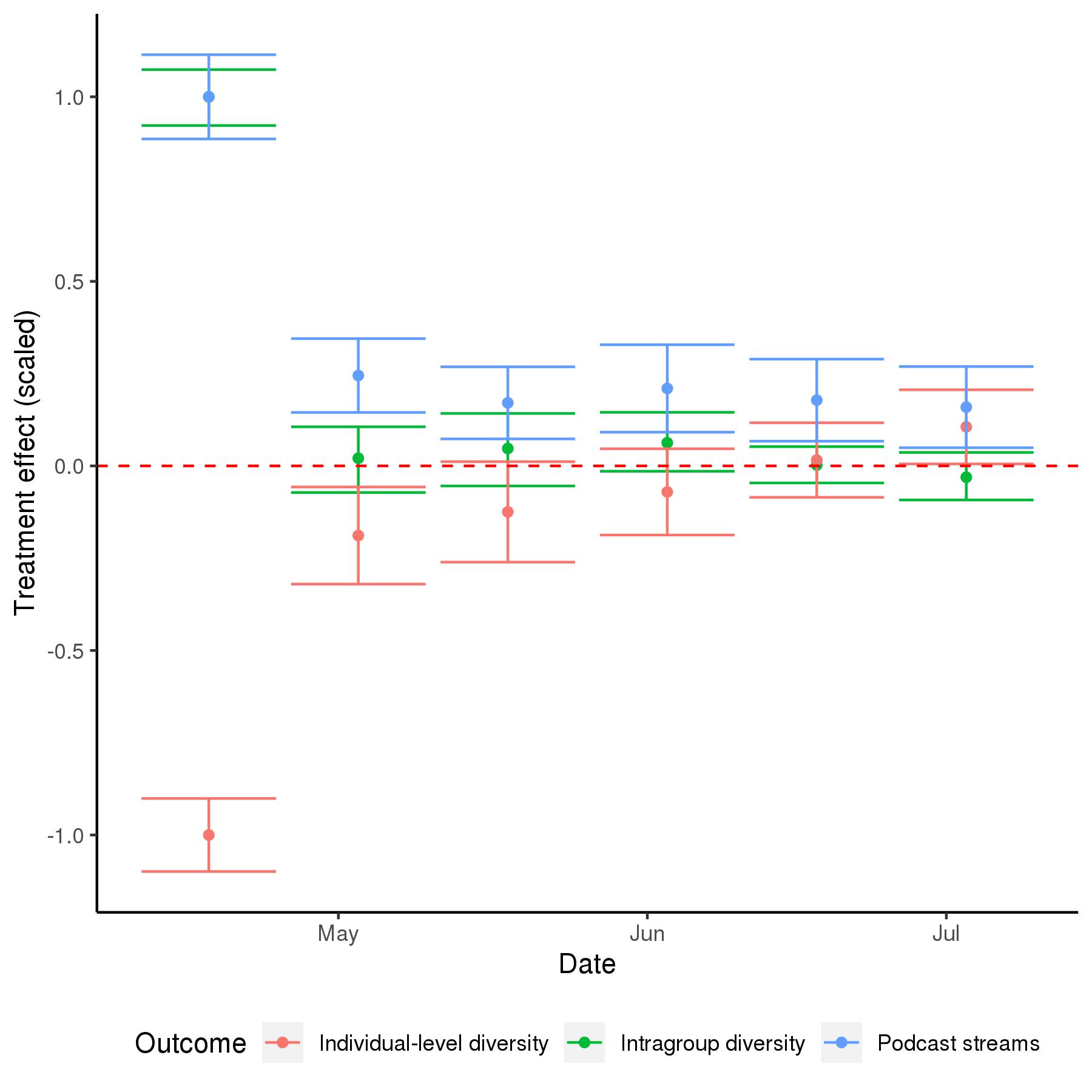}
\caption{The long-term effect of the treatment on podcast streams per user, individual-level streaming diversity conditional on streaming at least one podcast, and intragroup streaming diversity. Each outcome's time series is scaled by the absolute value of the magnitude of the treatment effect during the experiment.}
\label{fig:compare_time_series_streams_plot_popularity}
\end{center}
\end{figure}

\begin{figure}[!htbp]
\begin{center}
\includegraphics{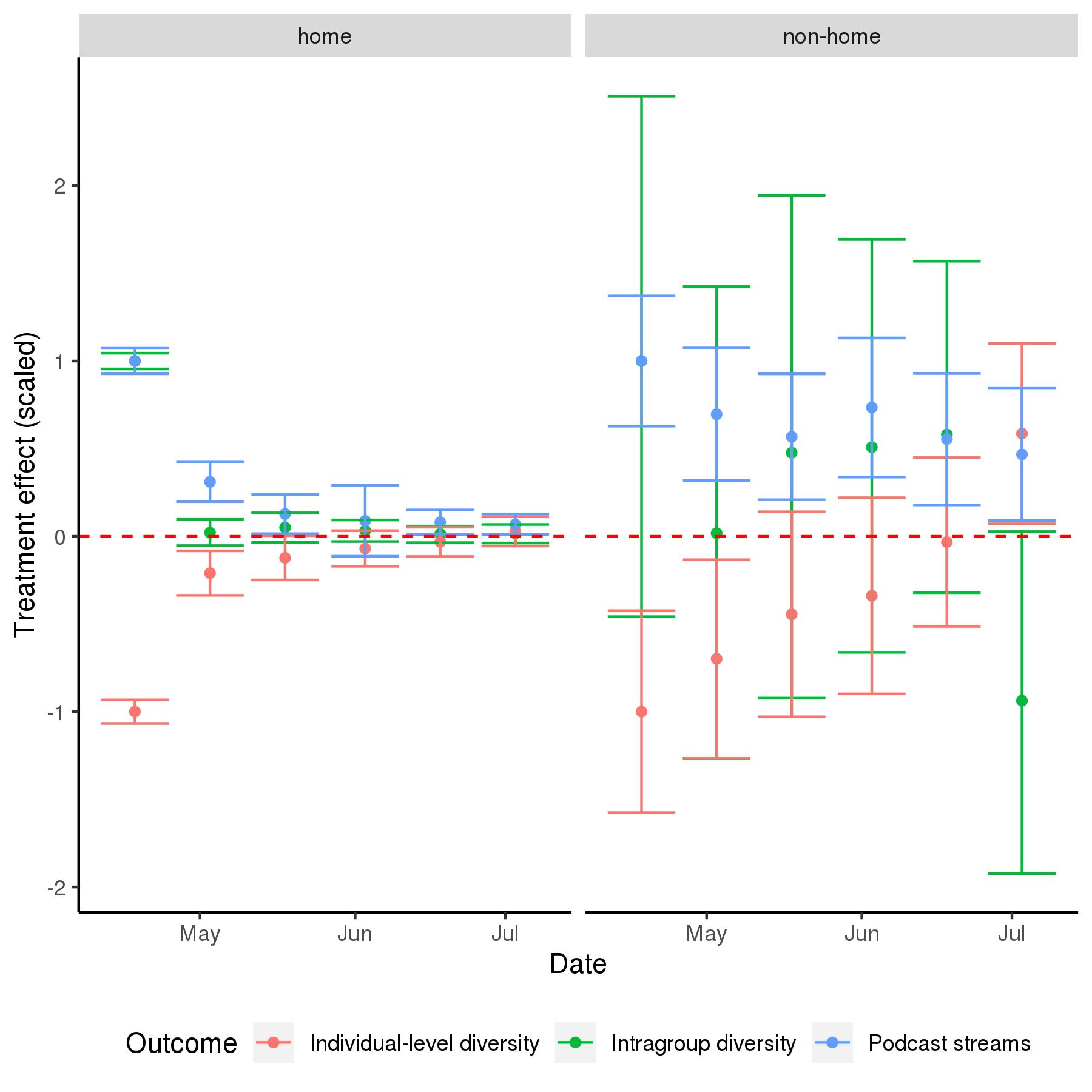}
\caption{The long-term referrer-level effect of the treatment on podcast streams per user, individual-level streaming diversity conditional on streaming at least one podcast, and intragroup streaming diversity. Each outcome's time series is scaled by the absolute value of the magnitude of the treatment effect during the experiment.}
\label{fig:compare_time_series_streams_referrer_plot_popularity}
\end{center}
\end{figure}

\begin{figure}[!htbp]
\begin{center}
\includegraphics[angle=-90, scale=.9]{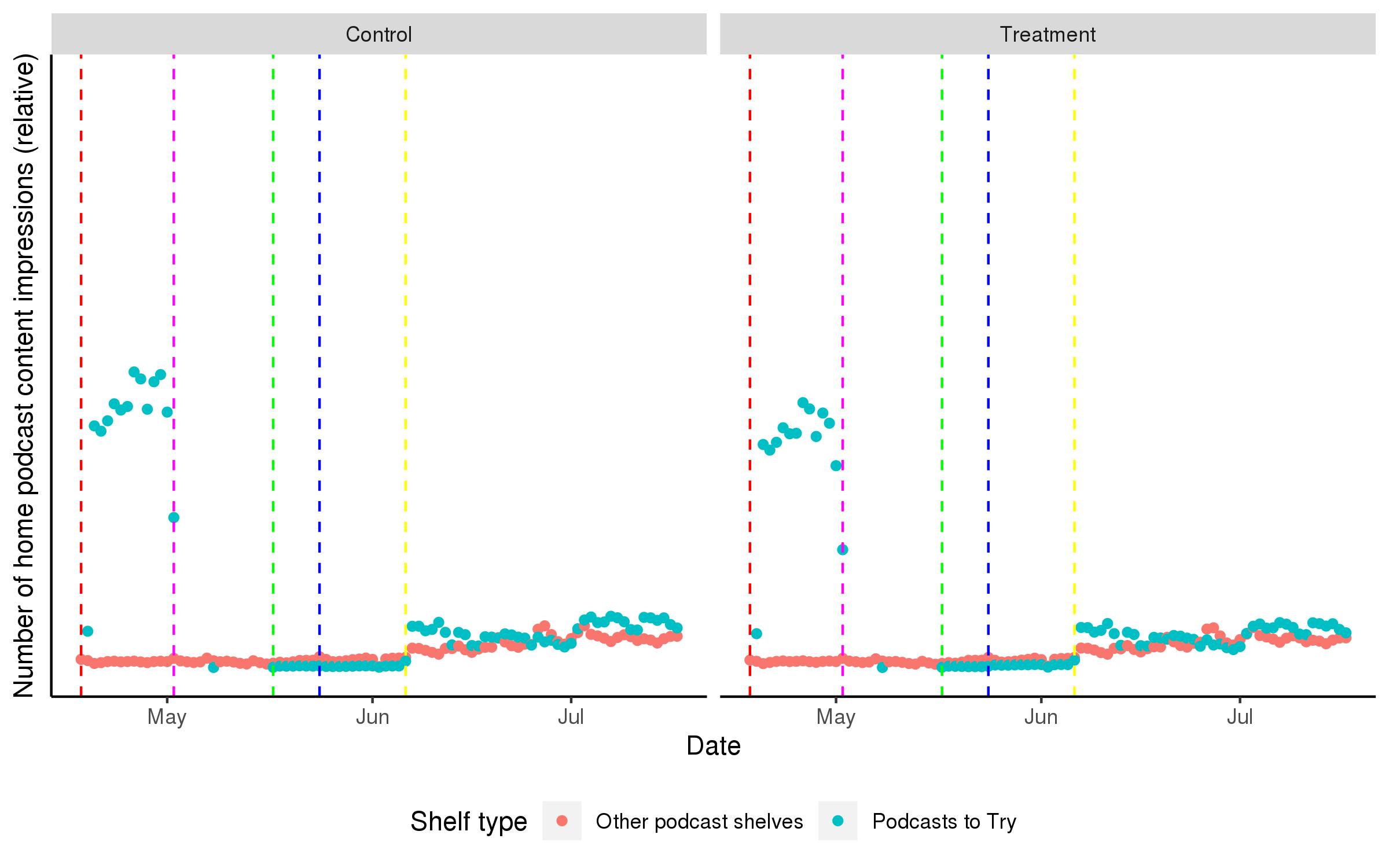}
\caption{The number of daily podcast content impressions from both the ``Podcasts to Try" shelf and other podcast-related shelves on the ``Home" section of the Spotify app, shown separately for users in the two treatment arms of the experiment. The dashed red line corresponds to the experiment launch date. The dashed magenta line corresponds to the experiment end date. The dashed green line corresponds to the productization of the ``Podcasts to try" shelf. The dashed blue line corresponds to the launch of the podcast shelf boosting experiment. The dashed yellow line corresponds to the end of the podcast shelf boosting experiment. y-axis values hidden due to confidentiality concerns.}
\label{fig:coldstart_shelf_impressions_simple_treatment}
\end{center}
\end{figure}

\begin{figure}[!htbp]
\begin{center}
\includegraphics{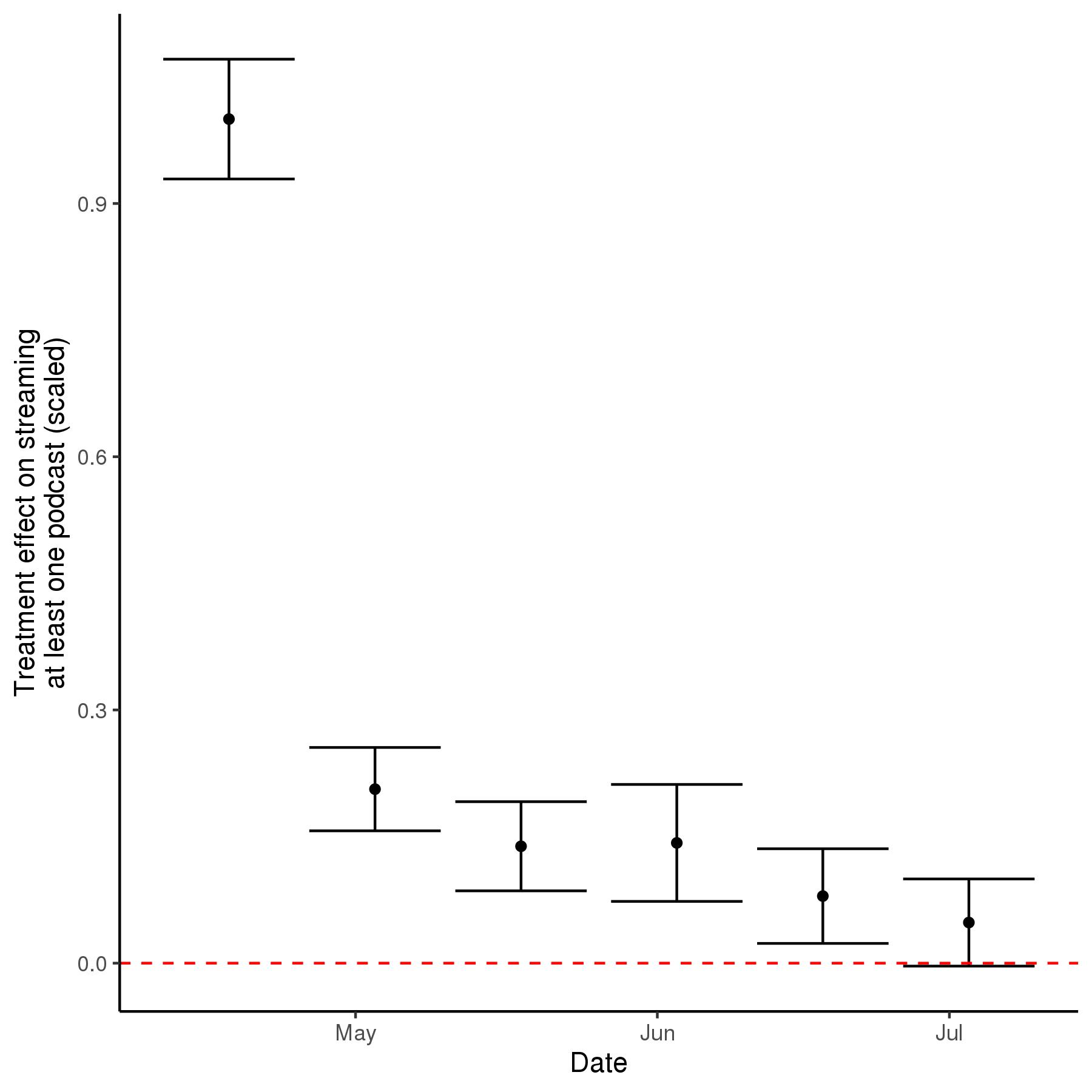}
\caption{The long-term effect of the treatment on the percentage of users streaming at least one podcast. The time series is scaled by the absolute value of the magnitude of the treatment effect during the experiment.}
\label{fig:users_streaming_dynamic}
\end{center}
\end{figure}

\begin{figure}[!htbp]
\begin{center} 
    \includegraphics{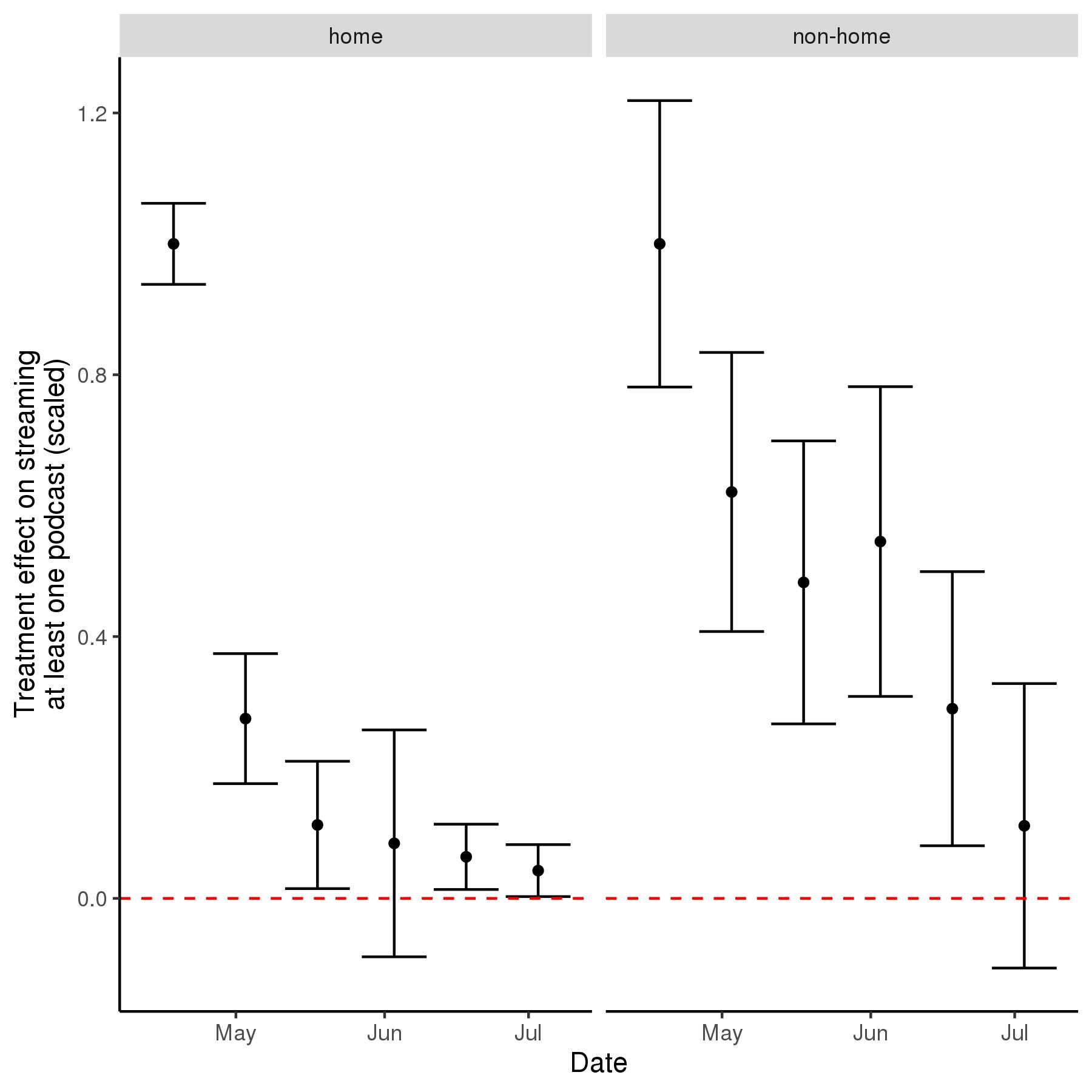}
    \caption{The long-term referrer-level effect of the treatment on the percentage of users streaming at least one podcast over time. Each time series is scaled by the absolute value of the magnitude of the treatment effect during the experiment.}
    \label{fig:user_level_streamers_surface_plot}
\end{center}
\end{figure}

\begin{figure}[!htbp]
\begin{center} 
    \includegraphics{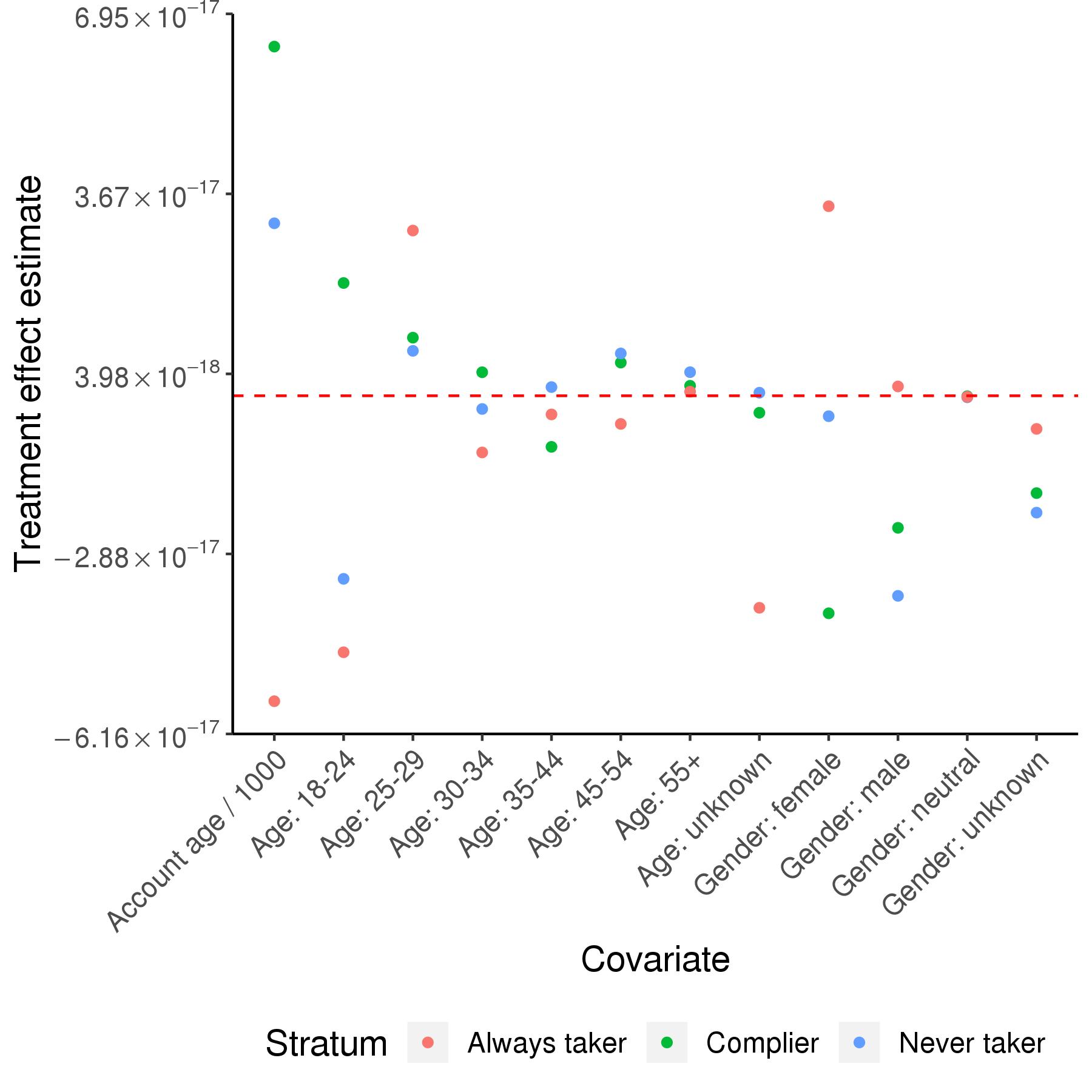}
    \caption{Results of the principal stratification balance check. The intermediate variable is whether a given user streamed at least one podcast during the experiment.}
    \label{fig:balance_checks_streams_plot_diversity_popularity}
\end{center}
\end{figure}

\begin{figure}[!htbp]
\begin{center} 
    \includegraphics{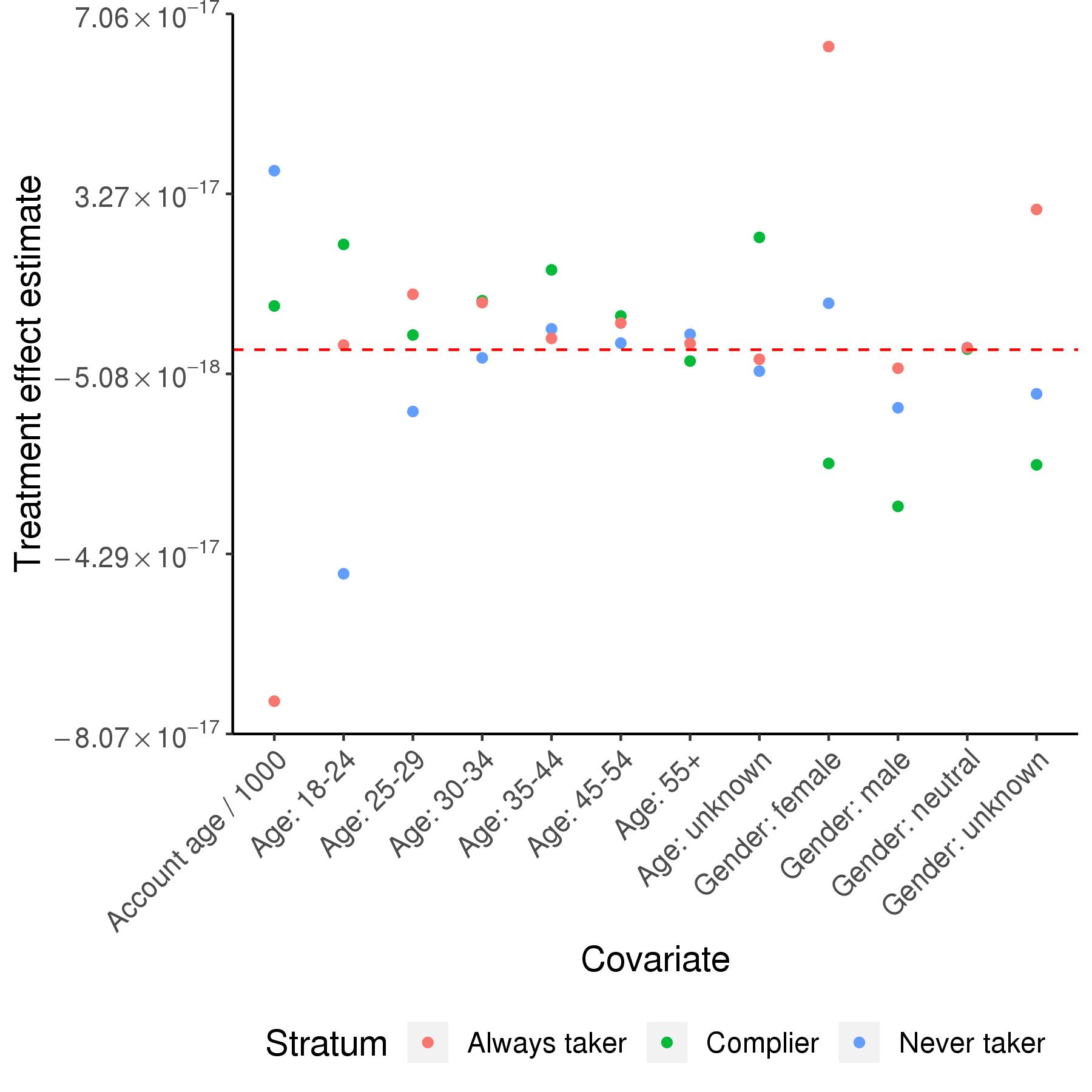}
    \caption{Results of the principal stratification balance check. The intermediate variable is whether a given user followed at least one podcast during the experiment.}
    \label{fig:balance_checks_follows_plot_diversity_popularity}
\end{center}
\end{figure}

\begin{figure}[!htbp]
\begin{center} 
    \includegraphics{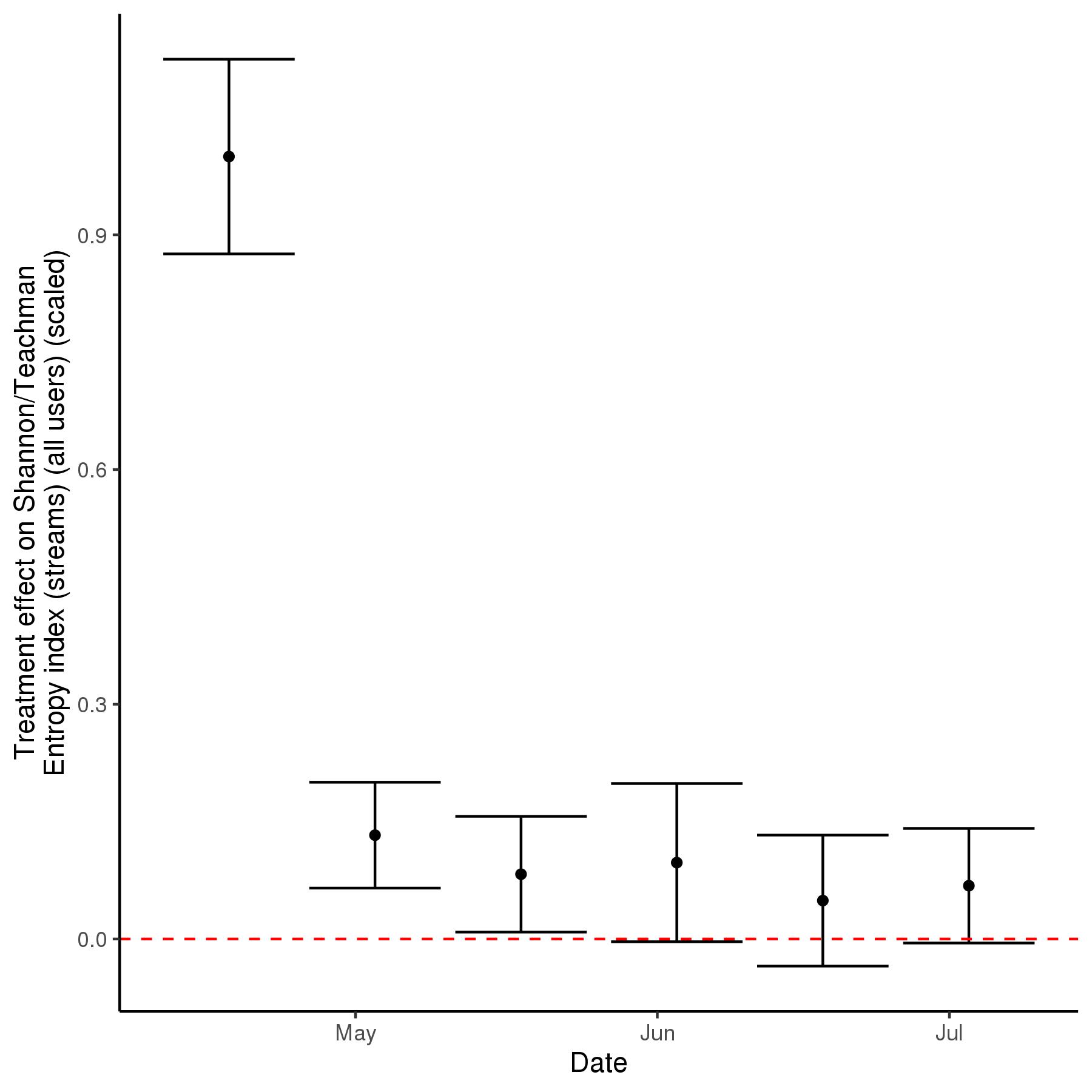}
    \caption{The long-term effect of the treatment on the average user-level Shannon entropy for streams. The time series is scaled by the absolute value of the magnitude of the treatment effect during the experiment.}
    \label{fig:individual_diversity_streams}
\end{center}
\end{figure}

\begin{figure}[!htbp]
\begin{center} 
\includegraphics{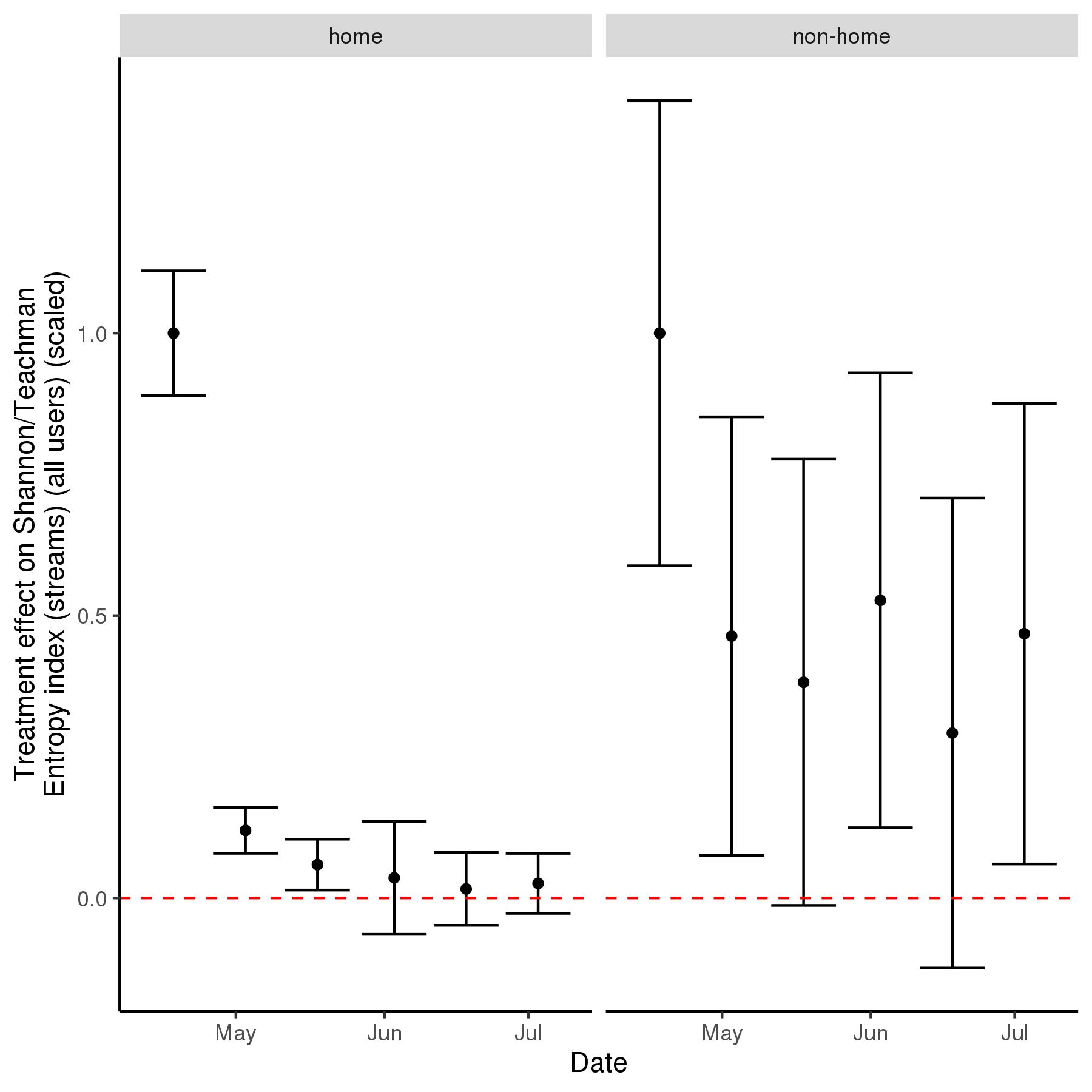}
\caption{The long-term referrer-level effect of the treatment on the average user-level Shannon entropy for streams over time. Each time series is scaled by the absolute value of the magnitude of the treatment effectduring the experiment.}
\label{fig:diversity_streamers_all_surface}
\end{center}
\end{figure}

\begin{figure}[!htbp]
\begin{center} 
\includegraphics{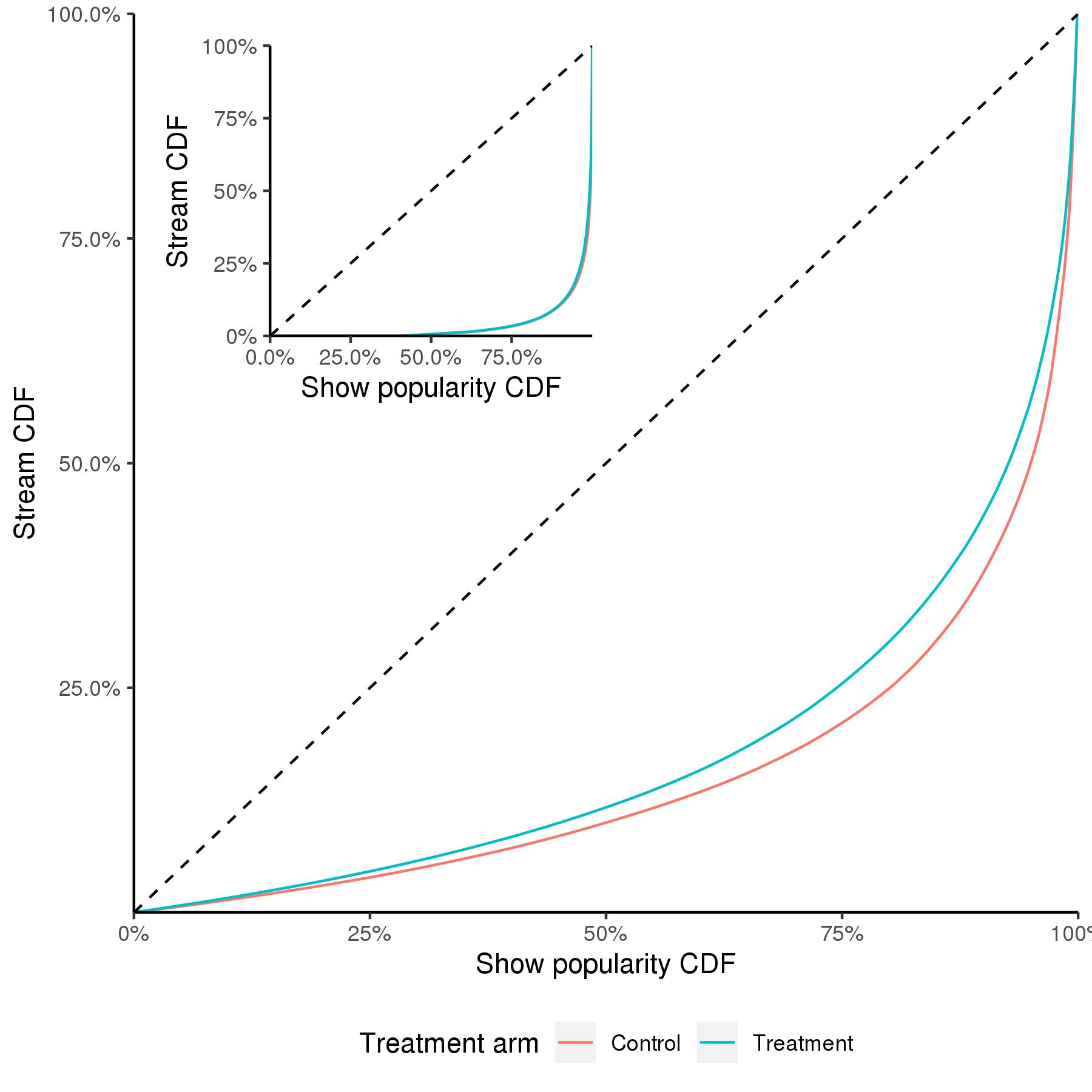}
\caption{The Lorenz curves for podcast streams, calculated separately for users in the treatment and control. The data for each Lorenz curve is limited to the 1,000 most streamed podcasts in the corresponding treatment arm data. The inset curve shows the Lorenz curve for streams across all podcasts.}
\label{fig:lorenz_streams}
\end{center}
\end{figure}

\begin{figure}[!htbp]
\begin{center} 
\includegraphics{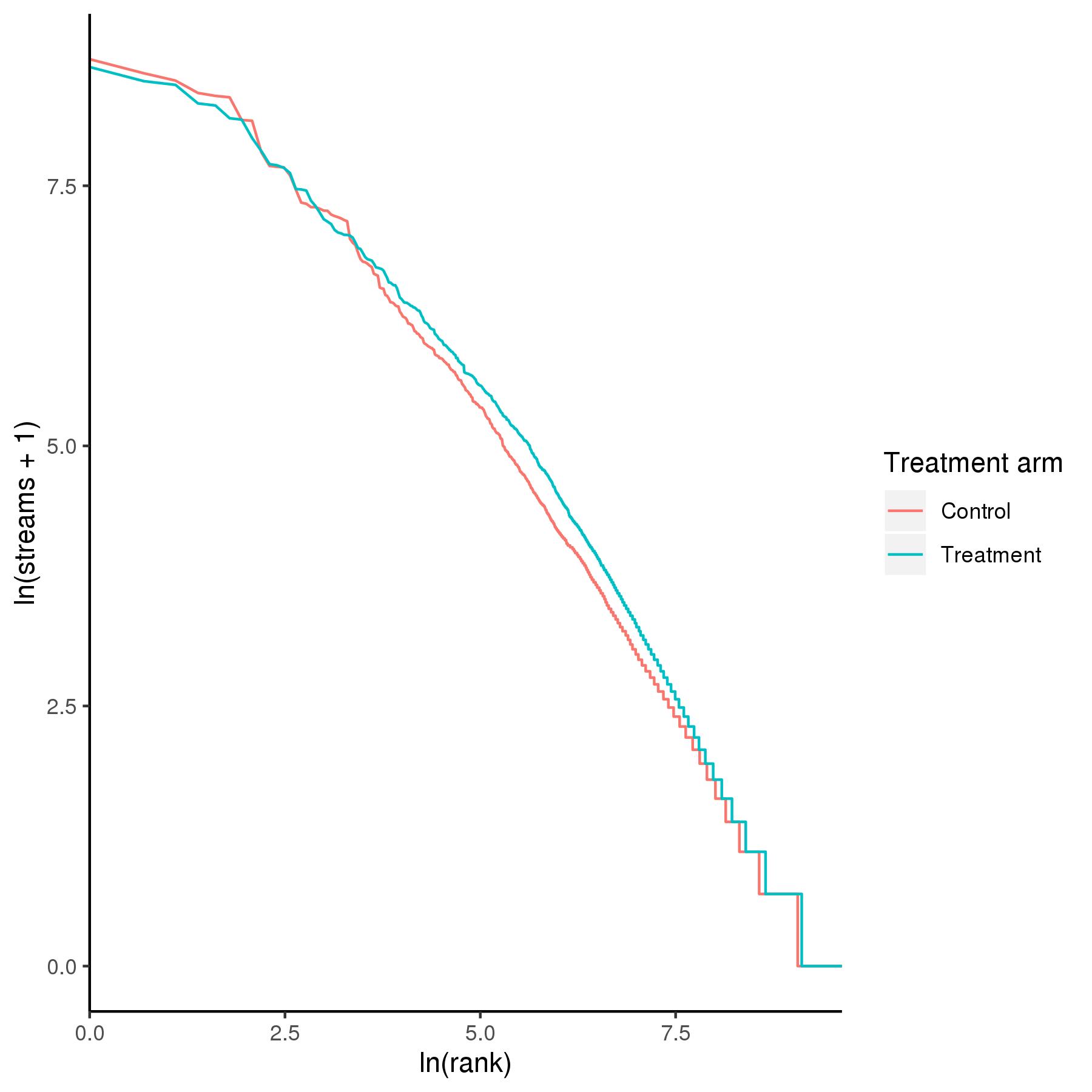}
\caption{The relationship between ln(streams + 1) and ln(stream rank) for both the control and treatment arms of the experiment.}
\label{fig:ln_ln_plot_immedate}
\end{center}
\end{figure}
 
\clearpage
\renewcommand\thetable{\thesection.\arabic{table}}    
\section{Additional Tables}
\setcounter{table}{0}    

\input{summary_stats_popularity.tex}

\input{shows_followed_model_popularity.tex}

\input{followed_shows_model_popularity.tex}

\input{individual_diversity_model_followers_only_popularity_no_obs.tex}

\input{individual_diversity_model_all_follows_popularity.tex}

\input{longtail_model_follows_immediate_popularity.tex}

\input{played_shows_model_popularity.tex}

\input{played_shows_referrer_popularity.tex}

\input{individual_diversity_model_all_popularity.tex}

\input{individual_diversity_model_all_streams_referrer_popularity.tex}

\input{longtail_model_streams_immediate_popularity.tex}

\end{APPENDICES}

\clearpage

\ACKNOWLEDGMENT{Analysis for this paper was conducted while David Holtz was an intern at Spotify during the summer of 2019. The authors are grateful to Samuel F. Way, Briana Vecchione, John Horton, Dean Eckles, Jui Ramaprasad, Joel Waldfogel, Daniel Rock, Michael Zhao, Katherine Hoffman Pham, Emma van Inwegen, Mazi Kazemi, Alex Moehring, Sebastian Steffen, Seth Benzell, Mahreen Khan, Hong Yi Tu Ye, Sanaz Mobasseri, and Martin Saveski for their helpful feedback. We also thank numerous other Spotify employees who have assisted with this project.}
%%TC:endignore

%%%%%%%%%%%%%%%%%
\end{document}

%% file: shows_played_model_popularity.tex
% Table created by stargazer v.5.2.2 by Marek Hlavac, Harvard University. E-mail: hlavac at fas.harvard.edu
% Date and time: Wed, Jan 15, 2020 - 02:50:54 PM
\begin{table}[!htbp] \centering 
  \caption{A linear model showing the effect of the 
treatment on number of podcasts streamed. Standard errors are clustered at the user bucket level.} 
  \label{tab:shows_played_model} 
\begin{tabular}{@{\extracolsep{5pt}}lcc} 
\\[-1.8ex]\hline 
\hline \\[-1.8ex] 
 & \multicolumn{2}{c}{\textit{Dependent variable:}} \\ 
\cline{2-3} 
\\[-1.8ex] & \multicolumn{2}{c}{Podcasts streamed} \\ 
\\[-1.8ex] & (1) & (2)\\ 
\hline \\[-1.8ex] 
 Treatment & 0.022$^{***}$ & 0.022$^{***}$ \\ 
  & (0.001) & (0.001) \\ 
  & & \\ 
 Constant & 0.077$^{***}$ & 0.056$^{***}$ \\ 
  & (0.001) & (0.001) \\ 
  & & \\ 
\hline \\[-1.8ex] 
User Gender & No & Yes \\ 
User Age & No & Yes \\ 
User account age & No & Yes \\ 
Observations & 852,937 & 852,937 \\ 
R$^{2}$ & 0.0005 & 0.003 \\ 
Adjusted R$^{2}$ & 0.0005 & 0.003 \\ 
Residual Std. Error & 0.508 (df = 852935) & 0.508 (df = 852925) \\ 
\hline 
\hline \\[-1.8ex] 
\textit{Note:}  & \multicolumn{2}{r}{$^{*}$p$<$0.1; $^{**}$p$<$0.05; $^{***}$p$<$0.01} \\ 
\end{tabular} 
\end{table}

%% file: individual_diversity_model_streamers_only_popularity_no_obs.tex
% Table created by stargazer v.5.2.2 by Marek Hlavac, Harvard University. E-mail: hlavac at fas.harvard.edu
% Date and time: Tue, Feb 11, 2020 - 06:06:57 PM
\begin{table}[!htbp] \centering 
  \caption{A linear model showing the difference in the average Shannon/Teachman entropy index (streams)
(podcast streamers only). Standard errors are clustered at the user bucket level.} 
  \label{tab:individual_diversity_model_streamers_only} 
\begin{tabular}{@{\extracolsep{5pt}}lcc} 
\\[-1.8ex]\hline 
\hline \\[-1.8ex] 
 & \multicolumn{2}{c}{\textit{Dependent variable:}} \\ 
\cline{2-3} 
\\[-1.8ex] & \multicolumn{2}{c}{Shannon/Teachman entropy index (streams)} \\ 
\\[-1.8ex] & (1) & (2)\\ 
\hline \\[-1.8ex] 
 Treatment & 0.071$^{***}$ & 0.070$^{***}$ \\ 
  & (0.004) & (0.004) \\ 
  & & \\ 
 Constant & 0.549$^{***}$ & 0.600$^{***}$ \\ 
  & (0.003) & (0.005) \\ 
  & & \\ 
\hline \\[-1.8ex] 
User Gender & No & Yes \\ 
User Age & No & Yes \\ 
User account age & No & Yes \\ 
R$^{2}$ & 0.005 & 0.016 \\ 
Adjusted R$^{2}$ & 0.005 & 0.016 \\ 
Residual Std. Error & 0.479 (df = 76191) & 0.477 (df = 76181) \\ 
\hline 
\hline \\[-1.8ex] 
\textit{Note:}  & \multicolumn{2}{r}{$^{*}$p$<$0.1; $^{**}$p$<$0.05; $^{***}$p$<$0.01} \\ 
 & \multicolumn{2}{r}{Observation counts hidden due to confidentiality concerns} \\ 
\end{tabular} 
\end{table}

%% file: shows_played_referrer_popularity.tex
% Table created by stargazer v.5.2.2 by Marek Hlavac, Harvard University. E-mail: hlavac at fas.harvard.edu
% Date and time: Wed, Jan 15, 2020 - 03:17:26 PM
\begin{table}[!htbp] \centering 
  \caption{A linear model showing the effect of the treatment on number of podcasts streamed, both on and off of home. Standard errors are clustered at the user bucket level.} 
  \label{tab:shows_played_model_referrer} 
\footnotesize 
\begin{tabular}{@{\extracolsep{5pt}}lcccc} 
\\[-1.8ex]\hline 
\hline \\[-1.8ex] 
 & \multicolumn{4}{c}{\textit{Dependent variable:}} \\ 
\cline{2-5} 
\\[-1.8ex] & \multicolumn{4}{c}{Podcasts streamed} \\ 
 & \multicolumn{2}{c}{Home} & \multicolumn{2}{c}{Non-home} \\ 
\\[-1.8ex] & (1) & (2) & (3) & (4)\\ 
\hline \\[-1.8ex] 
 Treatment & 0.020$^{***}$ & 0.020$^{***}$ & 0.006$^{***}$ & 0.006$^{***}$ \\ 
  & (0.001) & (0.001) & (0.001) & (0.001) \\ 
  & & & & \\ 
 Constant & 0.036$^{***}$ & 0.027$^{***}$ & 0.053$^{***}$ & 0.038$^{***}$ \\ 
  & (0.0005) & (0.001) & (0.001) & (0.001) \\ 
  & & & & \\ 
\hline \\[-1.8ex] 
User Gender & No & Yes & No & Yes \\ 
User Age & No & Yes & No & Yes \\ 
User account age & No & Yes & No & Yes \\ 
Observations & 852,937 & 852,937 & 852,937 & 852,937 \\ 
R$^{2}$ & 0.001 & 0.003 & 0.00004 & 0.002 \\ 
Adjusted R$^{2}$ & 0.001 & 0.003 & 0.00004 & 0.002 \\ 
Residual Std. Error & 0.260 (df = 852935) & 0.259 (df = 852925) & 0.447 (df = 852935) & 0.446 (df = 852925) \\ 
\hline 
\hline \\[-1.8ex] 
\textit{Note:}  & \multicolumn{4}{r}{$^{*}$p$<$0.1; $^{**}$p$<$0.05; $^{***}$p$<$0.01} \\ 
\end{tabular} 
\end{table}

%% file: individual_diversity_model_streams_only_referrer_popularity_no_obs.tex
% Table created by stargazer v.5.2.2 by Marek Hlavac, Harvard University. E-mail: hlavac at fas.harvard.edu
% Date and time: Tue, Feb 11, 2020 - 06:09:18 PM
\begin{table}[!htbp] \centering 
  \caption{A linear model showing the difference in the average Shannon/Teachman entropy 
index (streams) by stream referral source (podcast streamers only). Standard errors are clustered at the user bucket level.} 
  \label{tab:individual_diversity_model_streams_only_referrer} 
\small 
\begin{tabular}{@{\extracolsep{5pt}}lcccc} 
\\[-1.8ex]\hline 
\hline \\[-1.8ex] 
 & \multicolumn{4}{c}{\textit{Dependent variable:}} \\ 
\cline{2-5} 
\\[-1.8ex] & \multicolumn{4}{c}{Shannon/Teachman entropy index (streams)} \\ 
 & \multicolumn{2}{c}{Home} & \multicolumn{2}{c}{Non-home} \\ 
\\[-1.8ex] & (1) & (2) & (3) & (4)\\ 
\hline \\[-1.8ex] 
 Treatment & $-$0.116$^{***}$ & $-$0.116$^{***}$ & $-$0.018$^{***}$ & $-$0.017$^{***}$ \\ 
  & (0.004) & (0.004) & (0.005) & (0.005) \\ 
  & & & & \\ 
 Constant & 0.654$^{***}$ & 0.730$^{***}$ & 0.552$^{***}$ & 0.562$^{***}$ \\ 
  & (0.003) & (0.006) & (0.003) & (0.007) \\ 
  & & & & \\ 
\hline \\[-1.8ex] 
User Gender & No & Yes & No & Yes \\ 
User Age & No & Yes & No & Yes \\ 
User account age & No & Yes & No & Yes \\ 
R$^{2}$ & 0.016 & 0.029 & 0.0003 & 0.014 \\ 
Adjusted R$^{2}$ & 0.016 & 0.029 & 0.0003 & 0.014 \\ 
Residual Std. Error & 0.456 (df = 54335) & 0.453 (df = 54325) & 0.500 (df = 36327) & 0.497 (df = 36317) \\ 
\hline 
\hline \\[-1.8ex] 
\textit{Note:}  & \multicolumn{4}{r}{$^{*}$p$<$0.1; $^{**}$p$<$0.05; $^{***}$p$<$0.01} \\ 
 & \multicolumn{4}{r}{Observation counts hidden due to confidentiality concerns} \\ 
\end{tabular} 
\end{table}

%% file: summary_stats_popularity.tex
% latex table generated in R 3.6.1 by xtable 1.8-4 package
% Wed Jan 15 14:45:43 2020
\begin{table}[ht]
\centering
\caption{User bucket-level summary statistics for buckets in both the control and treatment arms of 
the experiment. $p$-values are computed using the Wilcoxon rank-sum test.} 
\label{tab:summary_stats}
\begingroup\scriptsize
\begin{tabular}{lllllll}
  \hline
Metric & Mean (control) & SD (control) & Mean (treatment) & SD (treatment) & $p$-value & Stat. sig \\ 
  \hline
Number of users & 4761.021 & 82.808 & 4713.965 & 89.633 & $<$ .001 & *** \\ 
  \% of users age 18 - 24 & 0.342 & 0.007 & 0.339 & 0.008 & 0.077 & * \\ 
  \% of users age 25 - 29 & 0.209 & 0.005 & 0.209 & 0.006 & 0.955 &  \\ 
  \% of users age 30 - 34 & 0.131 & 0.005 & 0.132 & 0.005 & 0.139 &  \\ 
  \% of users age 35 - 44 & 0.155 & 0.005 & 0.156 & 0.005 & 0.368 &  \\ 
  \% of users age 45 - 54 & 0.102 & 0.004 & 0.103 & 0.005 & 0.291 &  \\ 
  \% of users age 55+ & 0.055 & 0.003 & 0.055 & 0.003 & 0.542 &  \\ 
  \% of users of unknown age & 0.006 & 0.001 & 0.006 & 0.001 & 0.4 &  \\ 
  \% of male users & 0.537 & 0.006 & 0.538 & 0.007 & 0.322 &  \\ 
  \% of female users & 0.454 & 0.006 & 0.453 & 0.007 & 0.291 &  \\ 
  \% of users with other gender & 0.005 & 0.001 & 0.005 & 0.001 & 0.409 &  \\ 
  \% of users with gender unknown & 0.004 & 0.001 & 0.004 & 0.001 & 0.967 &  \\ 
  Average mean account age (days) & 1285.698 & 11.569 & 1284.711 & 11.435 & 0.412 &  \\ 
   \hline
\end{tabular}
\endgroup
\end{table}

%% file: shows_followed_model_popularity.tex
% Table created by stargazer v.5.2.2 by Marek Hlavac, Harvard University. E-mail: hlavac at fas.harvard.edu
% Date and time: Wed, Jan 15, 2020 - 03:00:52 PM
\begin{table}[!htbp] \centering 
  \caption{A linear model showing the effect of the treatment on number of podcasts followed. 
Standard errors are clustered at the user bucket level.} 
  \label{tab:shows_followed_model} 
\begin{tabular}{@{\extracolsep{5pt}}lcc} 
\\[-1.8ex]\hline 
\hline \\[-1.8ex] 
 & \multicolumn{2}{c}{\textit{Dependent variable:}} \\ 
\cline{2-3} 
\\[-1.8ex] & \multicolumn{2}{c}{Podcasts followed} \\ 
\\[-1.8ex] & (1) & (2)\\ 
\hline \\[-1.8ex] 
 Treatment & 0.012$^{***}$ & 0.012$^{***}$ \\ 
  & (0.001) & (0.001) \\ 
  & & \\ 
 Constant & 0.023$^{***}$ & 0.029$^{***}$ \\ 
  & (0.0005) & (0.001) \\ 
  & & \\ 
\hline \\[-1.8ex] 
User Gender & No & Yes \\ 
User Age & No & Yes \\ 
User account age & No & Yes \\ 
Observations & 852,937 & 852,937 \\ 
R$^{2}$ & 0.0004 & 0.001 \\ 
Adjusted R$^{2}$ & 0.0004 & 0.001 \\ 
Residual Std. Error & 0.301 (df = 852935) & 0.301 (df = 852925) \\ 
\hline 
\hline \\[-1.8ex] 
\textit{Note:}  & \multicolumn{2}{r}{$^{*}$p$<$0.1; $^{**}$p$<$0.05; $^{***}$p$<$0.01} \\ 
\end{tabular} 
\end{table}

%% file: followed_shows_model_popularity.tex
% Table created by stargazer v.5.2.2 by Marek Hlavac, Harvard University. E-mail: hlavac at fas.harvard.edu
% Date and time: Wed, Jan 15, 2020 - 03:00:54 PM
\begin{table}[!htbp] \centering 
  \caption{A linear probability model showing the effect of the treatment on following at least one podcast. 
Standard errors are clustered at the user bucket level.} 
  \label{tab:followed_show_model} 
\begin{tabular}{@{\extracolsep{5pt}}lcc} 
\\[-1.8ex]\hline 
\hline \\[-1.8ex] 
 & \multicolumn{2}{c}{\textit{Dependent variable:}} \\ 
\cline{2-3} 
\\[-1.8ex] & \multicolumn{2}{c}{Followed podcast} \\ 
\\[-1.8ex] & (1) & (2)\\ 
\hline \\[-1.8ex] 
 Treatment & 0.008$^{***}$ & 0.008$^{***}$ \\ 
  & (0.0003) & (0.0003) \\ 
  & & \\ 
 Constant & 0.015$^{***}$ & 0.018$^{***}$ \\ 
  & (0.0002) & (0.0004) \\ 
  & & \\ 
\hline \\[-1.8ex] 
User Gender & No & Yes \\ 
User Age & No & Yes \\ 
User account age & No & Yes \\ 
Observations & 852,937 & 852,937 \\ 
R$^{2}$ & 0.001 & 0.002 \\ 
Adjusted R$^{2}$ & 0.001 & 0.002 \\ 
Residual Std. Error & 0.135 (df = 852935) & 0.135 (df = 852925) \\ 
\hline 
\hline \\[-1.8ex] 
\textit{Note:}  & \multicolumn{2}{r}{$^{*}$p$<$0.1; $^{**}$p$<$0.05; $^{***}$p$<$0.01} \\ 
\end{tabular} 
\end{table}

%% file: individual_diversity_model_followers_only_popularity_no_obs.tex
% Table created by stargazer v.5.2.2 by Marek Hlavac, Harvard University. E-mail: hlavac at fas.harvard.edu
% Date and time: Tue, Feb 11, 2020 - 07:10:22 PM
\begin{table}[!htbp] \centering 
  \caption{A linear model showing the difference in the average Shannon/Teachman entropy index (follows) 
(podcast followers only). Standard errors are clustered at the user bucket level.} 
  \label{tab:individual_diversity_model_followers_only} 
\begin{tabular}{@{\extracolsep{5pt}}lcc} 
\\[-1.8ex]\hline 
\hline \\[-1.8ex] 
 & \multicolumn{2}{c}{\textit{Dependent variable:}} \\ 
\cline{2-3} 
\\[-1.8ex] & \multicolumn{2}{c}{Shannon/Teachman entropy index (follows)} \\ 
\\[-1.8ex] & (1) & (2)\\ 
\hline \\[-1.8ex] 
 Treatment & $-$0.069$^{***}$ & $-$0.068$^{***}$ \\ 
  & (0.008) & (0.008) \\ 
  & & \\ 
 Constant & 0.650$^{***}$ & 0.708$^{***}$ \\ 
  & (0.006) & (0.011) \\ 
  & & \\ 
\hline \\[-1.8ex] 
User Gender & No & Yes \\ 
User Age & No & Yes \\ 
User account age & No & Yes \\ 
R$^{2}$ & 0.005 & 0.021 \\ 
Adjusted R$^{2}$ & 0.005 & 0.020 \\ 
Residual Std. Error & 0.505 (df = 15894) & 0.501 (df = 15884) \\ 
\hline 
\hline \\[-1.8ex] 
\textit{Note:}  & \multicolumn{2}{r}{$^{*}$p$<$0.1; $^{**}$p$<$0.05; $^{***}$p$<$0.01} \\ 
 & \multicolumn{2}{r}{Observation counts hidden due to confidentiality concerns} \\ 
\end{tabular} 
\end{table}

%% file: individual_diversity_model_all_follows_popularity.tex
% Table created by stargazer v.5.2.2 by Marek Hlavac, Harvard University. E-mail: hlavac at fas.harvard.edu
% Date and time: Wed, Jan 15, 2020 - 08:39:02 PM
\begin{table}[!htbp] \centering 
  \caption{A linear model showing the effect of the treatment on the Shannon/Teachman entropy index (follows) 
(all users). Standard errors are clustered at the user bucket level.} 
  \label{tab:individual_diversity_model_all_follows} 
\begin{tabular}{@{\extracolsep{5pt}}lcc} 
\\[-1.8ex]\hline 
\hline \\[-1.8ex] 
 & \multicolumn{2}{c}{\textit{Dependent variable:}} \\ 
\cline{2-3} 
\\[-1.8ex] & \multicolumn{2}{c}{Shannon/Teachman entropy index (follows)} \\ 
\\[-1.8ex] & (1) & (2)\\ 
\hline \\[-1.8ex] 
 Treatment & 0.004$^{***}$ & 0.004$^{***}$ \\ 
  & (0.0003) & (0.0003) \\ 
  & & \\ 
 Constant & 0.010$^{***}$ & 0.013$^{***}$ \\ 
  & (0.0002) & (0.0003) \\ 
  & & \\ 
\hline \\[-1.8ex] 
User Gender & No & Yes \\ 
User Age & No & Yes \\ 
User account age & No & Yes \\ 
Observations & 852,937 & 852,937 \\ 
R$^{2}$ & 0.0003 & 0.001 \\ 
Adjusted R$^{2}$ & 0.0003 & 0.001 \\ 
Residual Std. Error & 0.108 (df = 852935) & 0.108 (df = 852925) \\ 
\hline 
\hline \\[-1.8ex] 
\textit{Note:}  & \multicolumn{2}{r}{$^{*}$p$<$0.1; $^{**}$p$<$0.05; $^{***}$p$<$0.01} \\ 
\end{tabular} 
\end{table}

%% file: longtail_model_follows_immediate_popularity.tex
% Table created by stargazer v.5.2.2 by Marek Hlavac, Harvard University. E-mail: hlavac at fas.harvard.edu
% Date and time: Tue, Jan 28, 2020 - 07:11:50 PM
\begin{table}[!htbp] \centering 
  \caption{Estimated coefficients for a model comparing the podcast follow Lorenz curves for control and treatment users} 
  \label{tab:ln_rank_model_follows} 
\begin{tabular}{@{\extracolsep{5pt}}lc} 
\\[-1.8ex]\hline 
\hline \\[-1.8ex] 
 & \multicolumn{1}{c}{\textit{Dependent variable:}} \\ 
\cline{2-2} 
\\[-1.8ex] & ln(follows + 1) \\ 
\hline \\[-1.8ex] 
 ln(rank) & $-$0.906$^{***}$ \\ 
  & ($-$0.915, $-$0.869) \\ 
  & \\ 
 Treatment & $-$0.208$^{***}$ \\ 
  & ($-$0.295, $-$0.079) \\ 
  & \\ 
 ln(rank) $\times$ Treatment & 0.172$^{***}$ \\ 
  & (0.133, 0.194) \\ 
  & \\ 
 Constant & 6.843$^{***}$ \\ 
  & (6.732, 6.892) \\ 
  & \\ 
\hline \\[-1.8ex] 
Observations & 400 \\ 
R$^{2}$ & 0.983 \\ 
Adjusted R$^{2}$ & 0.983 \\ 
Residual Std. Error & 0.111 (df = 396) \\ 
\hline 
\hline \\[-1.8ex] 
\textit{Note:}  & \multicolumn{1}{r}{$^{*}$p$<$0.1; $^{**}$p$<$0.05; $^{***}$p$<$0.01} \\ 
\end{tabular} 
\end{table}

%% file: played_shows_model_popularity.tex
% Table created by stargazer v.5.2.2 by Marek Hlavac, Harvard University. E-mail: hlavac at fas.harvard.edu
% Date and time: Wed, Jan 15, 2020 - 02:50:56 PM
\begin{table}[!htbp] \centering 
  \caption{A linear probability model showing the effect of the 
treatment on streaming at least one podcast. Standard errors are clustered at the user bucket level.} 
  \label{tab:played_show_model} 
\begin{tabular}{@{\extracolsep{5pt}}lcc} 
\\[-1.8ex]\hline 
\hline \\[-1.8ex] 
 & \multicolumn{2}{c}{\textit{Dependent variable:}} \\ 
\cline{2-3} 
\\[-1.8ex] & \multicolumn{2}{c}{Streamed podcast} \\ 
\\[-1.8ex] & (1) & (2)\\ 
\hline \\[-1.8ex] 
 Treatment & 0.017$^{***}$ & 0.017$^{***}$ \\ 
  & (0.001) & (0.001) \\ 
  & & \\ 
 Constant & 0.048$^{***}$ & 0.039$^{***}$ \\ 
  & (0.0004) & (0.001) \\ 
  & & \\ 
\hline \\[-1.8ex] 
User Gender & No & Yes \\ 
User Age & No & Yes \\ 
User account age & No & Yes \\ 
Observations & 852,937 & 852,937 \\ 
R$^{2}$ & 0.001 & 0.004 \\ 
Adjusted R$^{2}$ & 0.001 & 0.004 \\ 
Residual Std. Error & 0.230 (df = 852935) & 0.230 (df = 852925) \\ 
\hline 
\hline \\[-1.8ex] 
\textit{Note:}  & \multicolumn{2}{r}{$^{*}$p$<$0.1; $^{**}$p$<$0.05; $^{***}$p$<$0.01} \\ 
\end{tabular} 
\end{table}

%% file: played_shows_referrer_popularity.tex
% Table created by stargazer v.5.2.2 by Marek Hlavac, Harvard University. E-mail: hlavac at fas.harvard.edu
% Date and time: Wed, Jan 15, 2020 - 03:17:29 PM
\begin{table}[!htbp] \centering 
  \caption{A linear model showing the effect of the treatment on streaming at least one podcast, both on and off of home. Standard errors are clustered at the user bucket level.} 
  \label{tab:played_shows_model_referrer} 
\footnotesize 
\begin{tabular}{@{\extracolsep{5pt}}lcccc} 
\\[-1.8ex]\hline 
\hline \\[-1.8ex] 
 & \multicolumn{4}{c}{\textit{Dependent variable:}} \\ 
\cline{2-5} 
\\[-1.8ex] & \multicolumn{4}{c}{Stremed podcast} \\ 
 & \multicolumn{2}{c}{Home} & \multicolumn{2}{c}{Non-home} \\ 
\\[-1.8ex] & (1) & (2) & (3) & (4)\\ 
\hline \\[-1.8ex] 
 Treatment & 0.017$^{***}$ & 0.017$^{***}$ & 0.004$^{***}$ & 0.004$^{***}$ \\ 
  & (0.001) & (0.001) & (0.0004) & (0.0004) \\ 
  & & & & \\ 
 Constant & 0.029$^{***}$ & 0.023$^{***}$ & 0.030$^{***}$ & 0.026$^{***}$ \\ 
  & (0.0003) & (0.001) & (0.0003) & (0.001) \\ 
  & & & & \\ 
\hline \\[-1.8ex] 
User Gender & No & Yes & No & Yes \\ 
User Age & No & Yes & No & Yes \\ 
User account age & No & Yes & No & Yes \\ 
Observations & 852,937 & 852,937 & 852,937 & 852,937 \\ 
R$^{2}$ & 0.002 & 0.003 & 0.0001 & 0.004 \\ 
Adjusted R$^{2}$ & 0.002 & 0.003 & 0.0001 & 0.004 \\ 
Residual Std. Error & 0.190 (df = 852935) & 0.190 (df = 852925) & 0.175 (df = 852935) & 0.175 (df = 852925) \\ 
\hline 
\hline \\[-1.8ex] 
\textit{Note:}  & \multicolumn{4}{r}{$^{*}$p$<$0.1; $^{**}$p$<$0.05; $^{***}$p$<$0.01} \\ 
\end{tabular} 
\end{table}

%% file: individual_diversity_model_all_popularity.tex
% Table created by stargazer v.5.2.2 by Marek Hlavac, Harvard University. E-mail: hlavac at fas.harvard.edu
% Date and time: Wed, Jan 15, 2020 - 08:34:13 PM
\begin{table}[!htbp] \centering 
  \caption{A linear model showing the effect of the treatment on the Shannon/Teachman entropy index (streams)
(all users). Standard errors are clustered at the user bucket level.} 
  \label{tab:individual_diversity_model_all} 
\begin{tabular}{@{\extracolsep{5pt}}lcc} 
\\[-1.8ex]\hline 
\hline \\[-1.8ex] 
 & \multicolumn{2}{c}{\textit{Dependent variable:}} \\ 
\cline{2-3} 
\\[-1.8ex] & \multicolumn{2}{c}{Shannon/Teachman entropy index (streams)} \\ 
\\[-1.8ex] & (1) & (2)\\ 
\hline \\[-1.8ex] 
 Treatment & 0.010$^{***}$ & 0.010$^{***}$ \\ 
  & (0.001) & (0.001) \\ 
  & & \\ 
 Constant & 0.047$^{***}$ & 0.051$^{***}$ \\ 
  & (0.0004) & (0.001) \\ 
  & & \\ 
\hline \\[-1.8ex] 
User Gender & No & Yes \\ 
User Age & No & Yes \\ 
User account age & No & Yes \\ 
Observations & 852,937 & 852,937 \\ 
R$^{2}$ & 0.001 & 0.003 \\ 
Adjusted R$^{2}$ & 0.001 & 0.003 \\ 
Residual Std. Error & 0.219 (df = 852935) & 0.219 (df = 852925) \\ 
\hline 
\hline \\[-1.8ex] 
\textit{Note:}  & \multicolumn{2}{r}{$^{*}$p$<$0.1; $^{**}$p$<$0.05; $^{***}$p$<$0.01} \\ 
\end{tabular} 
\end{table}

%% file: individual_diversity_model_all_streams_referrer_popularity.tex
% Table created by stargazer v.5.2.2 by Marek Hlavac, Harvard University. E-mail: hlavac at fas.harvard.edu
% Date and time: Wed, Jan 15, 2020 - 08:57:03 PM
\begin{table}[!htbp] \centering 
  \caption{A linear model showing the effect of the treatment on the Shannon/Teachman entropy 
index (streams) by stream referral source (all users). Standard errors are clustered at the user bucket level.} 
  \label{tab:individual_diversity_model_all_streams_referrer} 
\footnotesize 
\begin{tabular}{@{\extracolsep{5pt}}lcccc} 
\\[-1.8ex]\hline 
\hline \\[-1.8ex] 
 & \multicolumn{4}{c}{\textit{Dependent variable:}} \\ 
\cline{2-5} 
\\[-1.8ex] & \multicolumn{4}{c}{Shannon/Teachman entropy index (streams)} \\ 
 & \multicolumn{2}{c}{Home} & \multicolumn{2}{c}{Non-home} \\ 
\\[-1.8ex] & (1) & (2) & (3) & (4)\\ 
\hline \\[-1.8ex] 
 Treatment & 0.010$^{***}$ & 0.010$^{***}$ & 0.002$^{***}$ & 0.002$^{***}$ \\ 
  & (0.001) & (0.001) & (0.0004) & (0.0004) \\ 
  & & & & \\ 
 Constant & 0.033$^{***}$ & 0.038$^{***}$ & 0.022$^{***}$ & 0.021$^{***}$ \\ 
  & (0.0004) & (0.001) & (0.0002) & (0.0005) \\ 
  & & & & \\ 
\hline \\[-1.8ex] 
User Gender & No & Yes & No & Yes \\ 
User Age & No & Yes & No & Yes \\ 
User account age & No & Yes & No & Yes \\ 
Observations & 852,937 & 852,937 & 852,937 & 852,937 \\ 
R$^{2}$ & 0.001 & 0.002 & 0.00003 & 0.002 \\ 
Adjusted R$^{2}$ & 0.001 & 0.002 & 0.00003 & 0.002 \\ 
Residual Std. Error & 0.184 (df = 852935) & 0.184 (df = 852925) & 0.151 (df = 852935) & 0.150 (df = 852925) \\ 
\hline 
\hline \\[-1.8ex] 
\textit{Note:}  & \multicolumn{4}{r}{$^{*}$p$<$0.1; $^{**}$p$<$0.05; $^{***}$p$<$0.01} \\ 
\end{tabular} 
\end{table}

%% file: longtail_model_streams_immediate_popularity.tex
% Table created by stargazer v.5.2.2 by Marek Hlavac, Harvard University. E-mail: hlavac at fas.harvard.edu
% Date and time: Tue, Jan 28, 2020 - 06:33:35 PM
\begin{table}[!htbp] \centering 
  \caption{Estimated coefficients for a model comparing the podcast stream Lorenz curves for control and treatment users} 
  \label{tab:ln_rank_model} 
\begin{tabular}{@{\extracolsep{5pt}}lc} 
\\[-1.8ex]\hline 
\hline \\[-1.8ex] 
 & \multicolumn{1}{c}{\textit{Dependent variable:}} \\ 
\cline{2-2} 
\\[-1.8ex] & ln(streams + 1) \\ 
\hline \\[-1.8ex] 
 ln(rank) & $-$1.046$^{***}$ \\ 
  & ($-$1.064, $-$1.024) \\ 
  & \\ 
 Treatment & $-$0.043 \\ 
  & ($-$0.178, 0.086) \\ 
  & \\ 
 ln(rank) $\times$ Treatment & 0.053$^{***}$ \\ 
  & (0.027, 0.078) \\ 
  & \\ 
 Constant & 10.453$^{***}$ \\ 
  & (10.380, 10.587) \\ 
  & \\ 
\hline \\[-1.8ex] 
Observations & 2,000 \\ 
R$^{2}$ & 0.987 \\ 
Adjusted R$^{2}$ & 0.987 \\ 
Residual Std. Error & 0.118 (df = 1996) \\ 
\hline 
\hline \\[-1.8ex] 
\textit{Note:}  & \multicolumn{1}{r}{$^{*}$p$<$0.1; $^{**}$p$<$0.05; $^{***}$p$<$0.01} \\ 
\end{tabular} 
\end{table}